\documentclass[11pt,a4paper]{scrartcl}
\usepackage{graphicx}
\usepackage{subcaption}
\usepackage{lcd}
\usepackage{lcd_title,ifthen}
\usepackage{mathptmx}
\usepackage{helvet}
\usepackage{amsmath}
\usepackage{color}
\usepackage{units}
\usepackage{url}
\usepackage{units}
\usepackage{heppennames2}
\usepackage{booktabs} 
\usepackage{float}
\usepackage{rccol}
\usepackage[numbers,sort]{natbib}
\usepackage[pdftex,bookmarks=true,breaklinks=true,bookmarksnumbered=true,colorlinks=true,linkcolor=webgreen,citecolor=webred,urlcolor=webblue]{hyperref}
\definecolor{webred}{rgb}{0.5, 0, 0}
\definecolor{webgreen}{rgb}{0, 0.5, 0}
\definecolor{webblue}{rgb}{0, 0, 0.5}
\makeatletter
\def\url@myurlfontstyle{%
    \@ifundefined{selectfont}{\def\UrlFont{\sf}}{\def\UrlFont{\small\ttfamily}}}
\makeatother
\long\def\symbolfootnote[#1]#2{\begingroup%
\def\thefootnote{\fnsymbol{footnote}}\footnote[#1]{#2}\endgroup} 
\lcdnote{2013-001}
\newlength{\capindent}
\newlength{\capwidth}
\newlength{\figwidth}
\newcommand{\icaption}[2][!*!,!]{\hspace*{\capindent}%
    \begin{minipage}{\capwidth}
        \ifthenelse{\equal{#1}{!*!,!}}%
            {\caption{#2}}%
            {\caption[#1]{#2}}
            \vspace*{3mm}
    \end{minipage}}
\newcommand{\abinv}{\ensuremath{\mathrm{ab}^{-1}}\xspace}
\newcommand{\smH}{\PH\xspace}
\newcommand{\ttH}{\ensuremath{\PQt\PAQt\smH}\xspace}
\newcommand{\toppair}{\ensuremath{\PQt\PAQt}\xspace}
\newcommand{\bpair}{\ensuremath{\PQb\PAQb}\xspace}
\newcommand{\gghadrons}{\mbox{\ensuremath{\PGg\PGg \rightarrow \mathrm{hadrons}}}\xspace}
\newcommand{\epluseminus}{\ensuremath{\Pep\Pem}\xspace}
\newcommand{\micron}{\ensuremath{\upmu\mathrm{m}}}

\begin{document}
\begin{titlepage}
%
\vskip 35mm
%
\mydocversion
%
\title{Measurement of the top Yukawa Coupling at a \unit[1]{TeV} International Linear Collider using the SiD detector}
%
\author{Philipp Roloff, Jan Strube}
\affiliations{CERN, CH-1211 Geneva 23, Switzerland}
%
\date{\today}
%
\begin{abstract}
\noindent
One of the detector benchmark processes investigated for the SiD Detailed Baseline Design (DBD) is given by: $\epluseminus \to \ttH$, where $\smH$ is the Standard Model Higgs boson of mass \unit[125]{GeV}. The study is carried out at a centre-of-mass energy of \unit[1]{TeV} and assuming an integrated luminosity of \unit[1]{ab$^{-1}$}. The physics aim is a direct measurement of the top Yukawa coupling at the ILC. Higgs boson decays to beauty quark-antiquark pairs are reconstructed. The investigated final states contain eight jets or six jets, one charged lepton and missing energy. Additionally, four of the jets in signal events are caused by beauty quark decays. The analysis is based on a full simulation of the SiD detector using \textsc{geant4}. Beam-related backgrounds from $\gamma\gamma \to \text{hadrons}$ interactions and incoherent $\epluseminus$ pairs are considered. This study addresses various aspects of the detector performance: jet clustering in complex hadronic final states, flavour-tagging and the identification of high energy leptons.
\end{abstract}

\end{titlepage}

\tableofcontents
  
\newpage

\section{Introduction}

The discovery of a Standard Model (SM)--like Higgs boson, announced on July 4th, 2012 
by the ATLAS and CMS collaborations~\cite{:2012gu,:2012gk},
was celebrated as a major milestone of modern physics. It represents the
start of an era in which the properties of this particle will be measured with
the best possible precision. 

The SM predicts a linear dependence between the
coupling strength of the Higgs boson to a fermion and the fermion mass. Since the
top quark is the heaviest known fundamental particle, the measurement of
the top Yukawa coupling serves as the high endpoint to test this prediction. In the SM, the top Yukawa coupling, $y_{t}$, 
has a value of:
\begin{equation}
y_{t} = \sqrt{2} \frac{m_{t}}{v},
\end{equation}
where $m_{t}$ is the top quark mass and $v$ is the Higgs vacuum expectation value. 
On the other hand, sizeable deviations of the top Yukawa coupling from the SM prediction 
are expected in various new physics scenarios~\cite{Gupta:2012mi}.

The International Linear Collider (ILC)~\cite{BrauJames:2007aa} is a proposed \epluseminus collider with a
maximum centre-of-mass energy $\sqrt{s} = \unit[1]{TeV}$. It has a broad physics potential that is complementary
to the LHC. Measurements of the Higgs couplings with the utmost precision
are an integral part of the physics programme at this machine.

In the following, a study of the physics potential for a measurement of the top Yukawa coupling at 
$\sqrt{s} = \unit[1]{TeV}$ using the \texttt{sidloi3} detector concept is presented. An integrated luminosity of \unit[1]{ab$^{-1}$} is assumed. The status of the analysis 
described in this note corresponds to the results given in the 
chapter on physics benchmark studies of the SiD DBD report~\cite{sid_dbd}. Future extensions of this 
study are foreseen.

The results reported in this document complement earlier 
feasibility studies~\cite{Hagiwara1991257,doi:10.1142/S0217732392001464,Juste:1999af,Gay:2006vs,PhysRevD.61.013002}.
Recently a study based on a fast Monte Carlo detector simulation was performed assuming a centre-of-mass energy of 
$\sqrt{s} = \unit[500]{\;GeV}$~\cite{PhysRevD.84.014033}.

\section{Properties of beam-induced backgrounds}
\label{sec:beam_backgrounds}
\begin{figure}
    \centering
    \includegraphics[width=.49\textwidth]{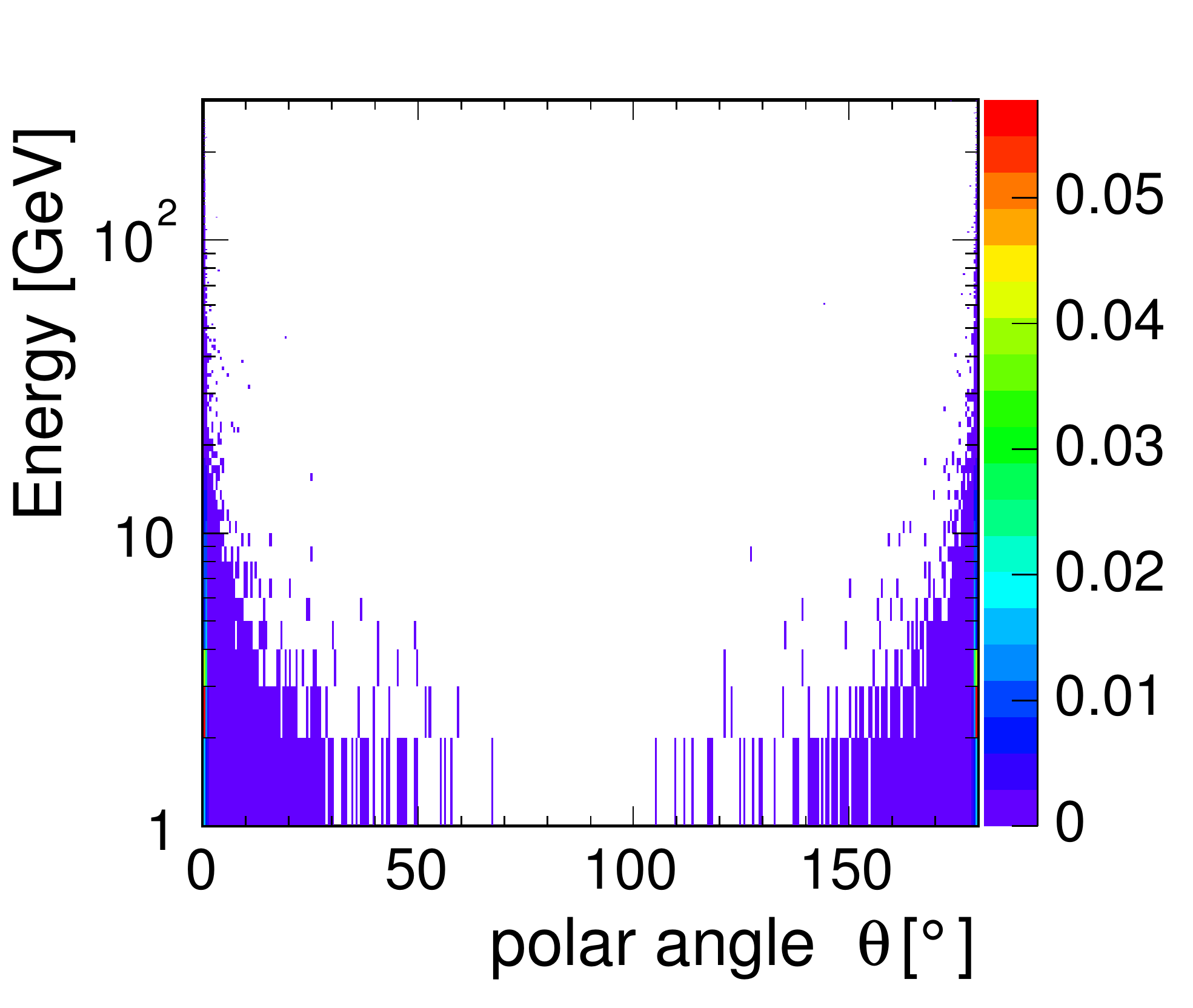}
    \includegraphics[width=.49\textwidth]{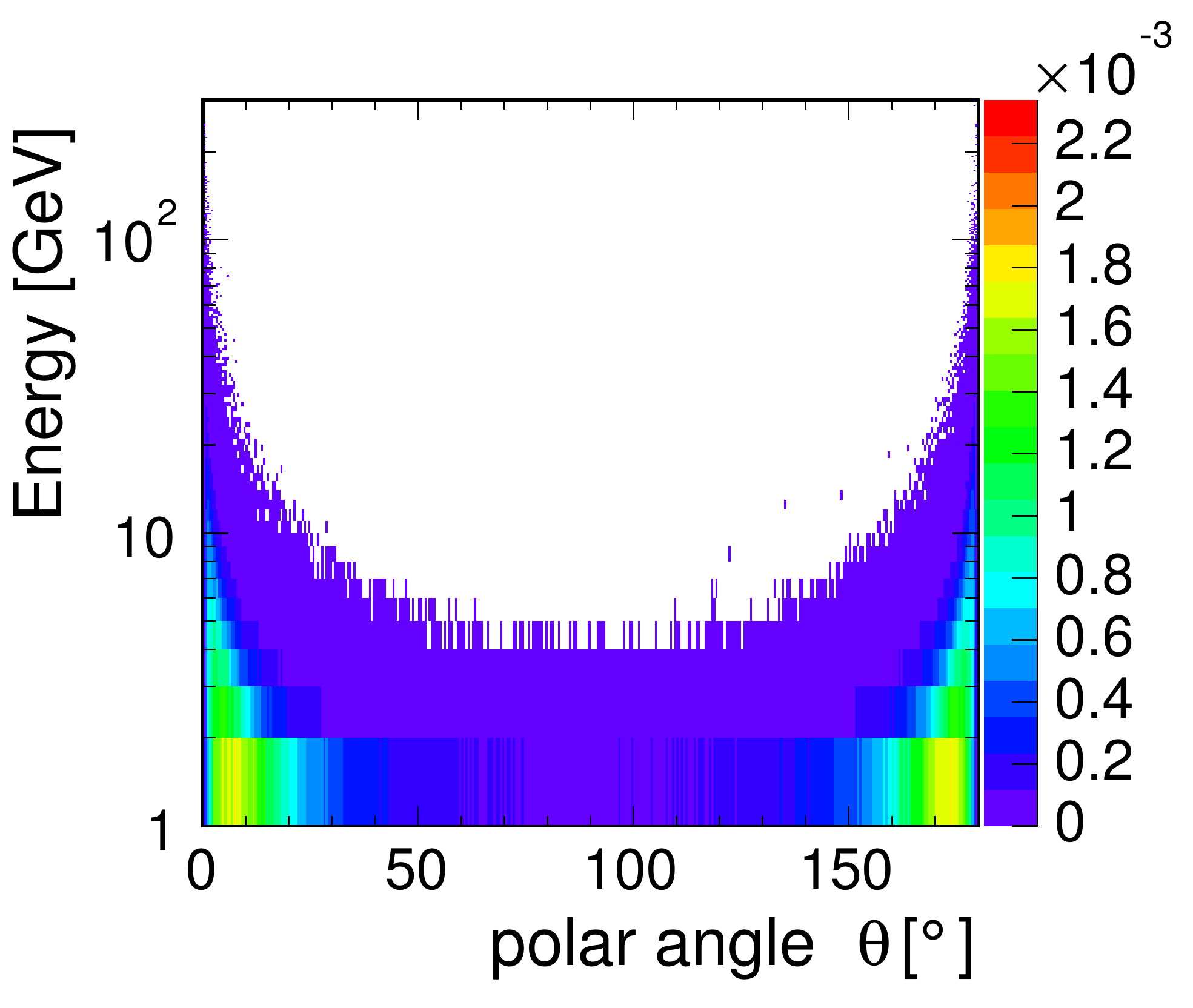}
    \caption{Energy Distribution of the simulated particles from incoherent pair production (left) and from \gghadrons processes (right). Both distributions are normalised to unity.}
    \label{fig:incoherentPairsGGHadMonteCarlo}
\end{figure}

The \unit[1]{TeV} ILC has an instantaneous luminosity of \unit[$4.2 \times 10^{34}$]{$\mathrm{cm}^{-2}\mathrm{s}^{-1}$}. During the collision a number of processes occur in addition
to the primary scattering event. For this analysis, two kinds of these processes
were simulated: The production of incoherent electron-positron pairs resulting in an average of 450000 low-momentum particles per bunch crossing and the 
production of hadronic final states from an average of 4.1 two-photon events per bunch crossing.
Figure~\ref{fig:incoherentPairsGGHadMonteCarlo} shows the distribution of the energy
of the simulated particles versus their polar angle~$\theta$. 

While the most energetic
particles from incoherent pair production are primarily outside of the detector acceptance,
some low-$p_\mathrm{T}$ particles lead to an occupancy of up to \unit[0.06]{$\text{hits} / \mathrm{mm}^{2} / \text{BX}$} in the vertex detector and up to \unit[$5\times10^{-5}$]{$\text{hits}/\mathrm{mm}^{2} / \text{BX}$} in the main tracker for the \texttt{sidloi3} detector model without the field from the DID magnet~\cite{sid_workshop_2013_grefe}. Particles from \gghadrons
processes on the other hand can have sizeable values of $p_\mathrm{T}$ and reach the calorimeters,
where they impact on the energy reconstruction of the primary physics process. The 
beam-induced backgrounds do not degrade the tracking performance significantly~\cite{sid_dbd}.

The primary vertices of these processes are distributed with a Gaussian profile along the beam direction across the luminous region of \unit[225]{\micron}.

\section{Analysis framework}
Top pair events were generated using the \textsc{whizard 1.95}~\cite{Kilian:2007gr, Moretti:2001zz} Monte Carlo generator while all other samples were obtained using \textsc{physsim}~\cite{physsim-homepage}. The expected luminosity spectrum at the ILC was taken into account during the event generation. The model for the non-perturbative hadronisation in \textsc{pythia 6.4}~\cite{Sjostrand2006} with a tune of the hadronisation parameters to \textsc{opal} data\footnote{The exact values of all paramteres are listed for example in Appendix B of~\cite{Linssen:2012hp}} was used for all samples.

All events are simulated in the \texttt{sidloi3} model of the SiD detector concept using \textsc{slic}~\cite{Graf:2006ei}, which is a thin wrapper around \textsc{geant4}~\cite{Agostinelli2003,Allison2006}. They are reconstructed by the algorithms in the \texttt{org.lcsim}~\cite{lcsim} and \texttt{slicPandora}~\cite{Thomson:2009rp} programs. The LCFIPlus~\cite{LCFIPlus} package is used for flavour tagging. The assumed integrated luminosity of the analysis is \unit[1]{\abinv}, which is split equally between the two polarisation states $(+80\%, -20\%)$ and $(-80\%, +20\%)$ for the polarization of electron and positron beams $(P_{\Pem}, P_{\Pep})$. Background from processes described in Section~\ref{sec:beam_backgrounds} is overlayed on the hit level before the digitization using a procedure~\cite{lcd-2011-032} originally developed for CLIC.

\section{The \texttt{sidloi3} detector model}
\label{sec:DetectorModel}
The \texttt{sidloi3} detector model, in which these studies are carried out, is a general-purpose detector with a \unit[4]{$\pi$} coverage as described in the SiD Letter of Intent~\cite{Aihara:2009ad}. It is designed for particle flow calorimetry using highly granular calorimeters.

A superconducting solenoid with an inner radius of \unit[2.6]{m} provides a central magnetic field of \unit[5]{T}. The calorimeters are placed inside the coil and consist of a 30 layer tungsten--silicon electromagnetic calorimeter with \unit[13]{$\mathrm{mm}^2$} segmentation, followed by a hadronic cal\-o\-rim\-e\-ter with steel absorber and instrumented with resistive plate chambers (RPC) -- 40 layers in the barrel region and 45 layers in the endcaps. The read-out cell size in the hadronic calorimeters is \unit[$10\times10$]{$\mathrm{mm}^2$}. The iron return yoke outside of the coil is instrumented with 11 RPC layers with \unit[$30\times30$]{$\mathrm{mm}^2$} read-out cells for muon identification.

The silicon-only tracking system consists of five \unit[$20\times20$]{$\micron^2$} pixel layers followed by five strip layers with a pitch of \unit[25]{\micron}, a read-out pitch of \unit[50]{\micron} and a length of \unit[92]{mm} in the barrel region. The tracking system in the endcap consists of four stereo-strip disks with similar pitch and a stereo angle of $12^\circ$, complemented by four pixelated disks in the vertex region with a pixel size of \unit[$20\times 20$]{$\micron^2$} and three disks in the far-forward region at lower radii with a pixel size of \unit[$50\times50$]{$\micron^2$}.
All sub-detectors have time-stamping capability that allow to separate hits originating from different bunch crossings.

\section{Analysis strategy and Monte Carlo samples}
\begin{figure}
\centering
\includegraphics[width=0.40\textwidth]{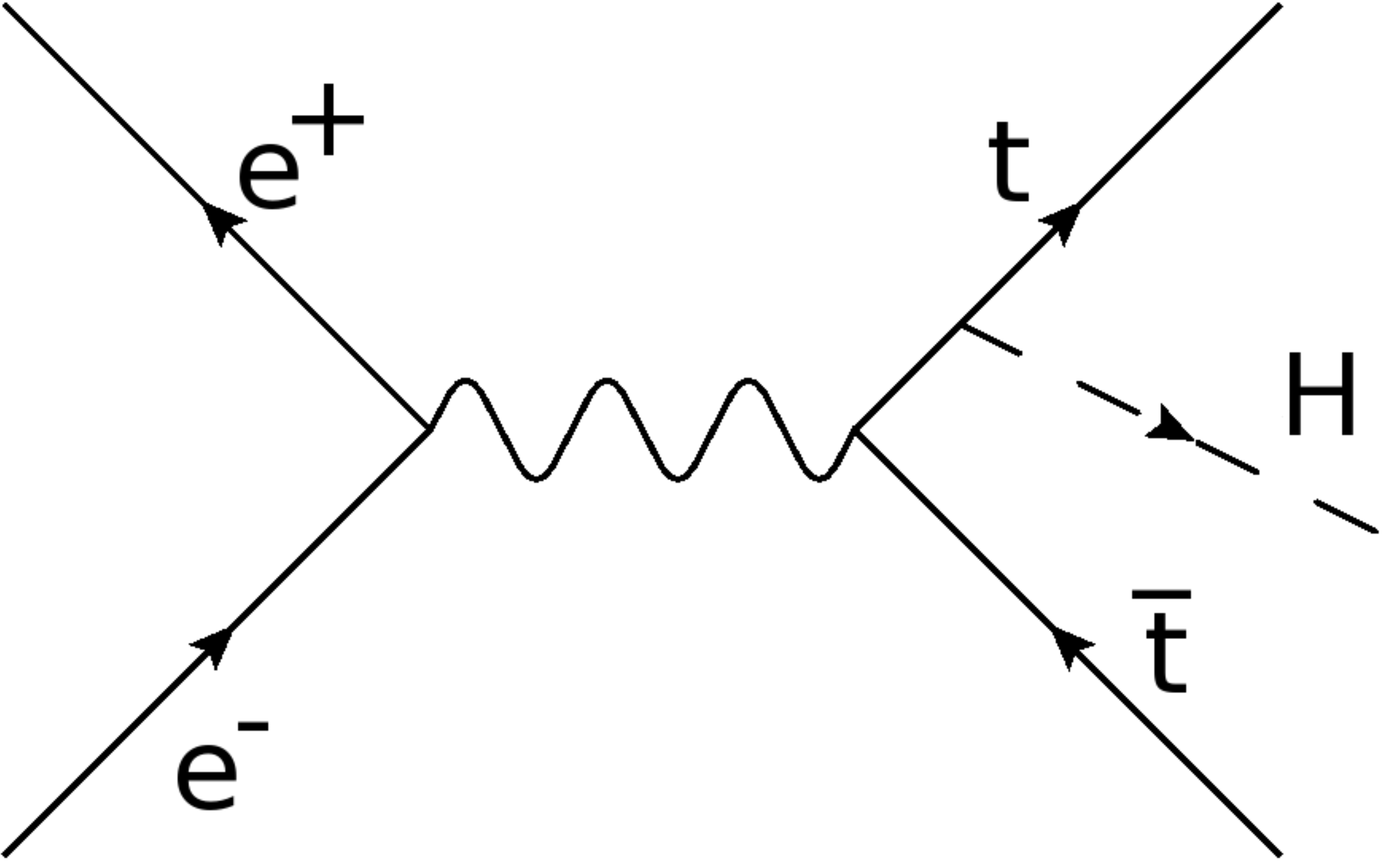} \hspace{2cm}
\includegraphics[width=0.37\textwidth]{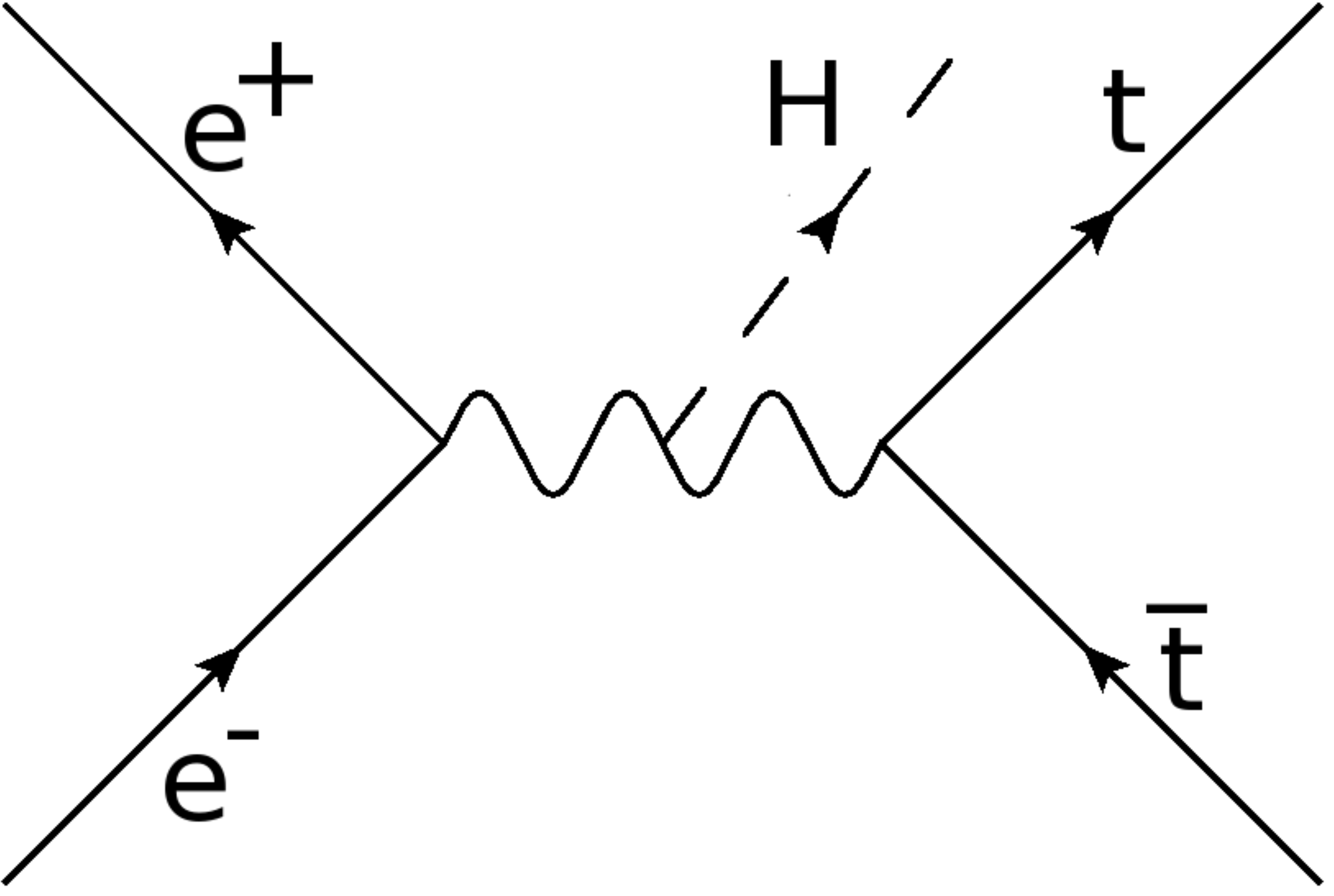}
\caption{\label{sid:benchmarking:fig:feynman_diagrams_tth} Diagrams for \ttH production in \epluseminus collisions.}
\end{figure}
The measurement of the cross section for the process $\epluseminus \to \ttH$  using two different final states is 
described in the following. The Feynman diagrams for this process are shown in
Figure~\ref{sid:benchmarking:fig:feynman_diagrams_tth}. Here $\smH$ is a
Standard Model Higgs boson of mass 125~GeV. The diagram shown on the left
represents the dominant contribution to the cross section. This diagram is 
directly sensitive to the top Yukawa coupling $y_{t}$. The contribution from 
Higgsstrahlung off the intermediate $\PZ$ boson increases the cross 
section for the \ttH final state by about 4\%~\cite{physsim-homepage}. This represents a
small correction which needs to be taken into account in the extraction of
$y_{t}$ from the measured cross section. The correction will be known with good precision, 
because the Higgs coupling to the \PZ boson can be extracted from measurements of 
$\epluseminus \to \smH\PZ$ events at $\sqrt{s} = \unit[250]{GeV}$ with a statistical unceratainty of 
about 1.5\%~\cite{Aihara:2009ad, Brau:2012hv}.

The measurement of the \ttH cross section at the ILC allows a direct extraction of
the top Yukawa coupling with good precision. In the analysis presented here, the
Higgs decay $\smH \to \bpair$ is considered. Two final states are
investigated in the following:
\begin{itemize}
\item{\textbf{8 jets:} In this case both \PW bosons from $\toppair$ decay hadronically. Hence this 
final state contains eight jets out of which four originate from b-quark decays.}
\item{\textbf{6 jets:} Here one \PW boson decays hadronically and the other \PW boson 
decays leptonically. The final state contains four b-jets, two further jets, 
an isolated lepton and missing energy. Only electrons and muons are considered as 
isolated leptons in the final state.}
\end{itemize}
This study requires jet clustering in complex hadronic final states, missing energy 
reconstruction, flavour-tagging and reconstruction and identification of high 
energy leptons. Hence it represents a comprehensive check of the complete analysis chain 
and overall detector performance.

\begin{table}
\caption{\label{sid:benchmarking:tab:tth_cross_sections} Production cross sections times 
branching ratios or production cross sections for the signal final states and for the considered backgrounds. All samples were 
generated assuming a Standard Model Higgs with a mass of \unit[125]{GeV}. The numbers for ``other \ttH''
processes in this table do not include either of the signal final states (see text). The $\toppair \PZ$ and $\toppair \Pg^{*}$ samples, 
where $\Pg^{*}$ is a hard gluon splitting into a $\bpair$ pair, 
do not contain events where both top quarks decay leptonically. The $\toppair$ samples contain all possible 
decays of both \PW bosons.}
\begin{center}
\begin{tabular}{l c c c c} \toprule
Type & Final state & P(\Pem) & P(\Pep) & Cross section [$\times$ BR] (fb) \\ \midrule
Signal & \ttH (8 jets) & $-80\%$ & $+20\%$ & 0.87 \\
Signal & \ttH (8 jets) & $+80\%$ & $-20\%$ & 0.44 \\
Signal & \ttH (6 jets) & $-80\%$ & $+20\%$ & 0.84 \\
Signal & \ttH (6 jets) & $+80\%$ & $-20\%$ & 0.42 \\ \midrule
Background & other \ttH & $-80\%$ & $+20\%$ & 1.59 \\
Background & other \ttH & $+80\%$ & $-20\%$ & 0.80 \\
Background & $\toppair Z$ & $-80\%$ & $+20\%$ & 6.92 \\
Background & $\toppair Z$ & $+80\%$ & $-20\%$ & 2.61 \\
Background & $\toppair g^{*} \to \toppair\bpair$ & $-80\%$ & $+20\%$ & 1.72 \\
Background & $\toppair g^{*} \to \toppair\bpair$ & $+80\%$ & $-20\%$ & 0.86 \\
Background & $\toppair$ & $-80\%$ & $+20\%$ & 449 \\
Background & $\toppair$ & $+80\%$ & $-20\%$ & 170 \\ \bottomrule
\end{tabular}
\end{center}
\end{table}

An overview of the cross sections for the signal final states as well as for the
considered backgrounds is shown in Table~\ref{sid:benchmarking:tab:tth_cross_sections}. 
For the measurement using the final state with six jets, all other $\ttH$ events, i.e., all 
events where both top quarks decay leptonically or hadronically, or events where the 
Higgs boson does not decay into $\bpair$, are treated as background. For the eight 
jets final state events where at least one top quark decays leptonically or where the Higgs 
boson does not decay into $\bpair$ are considered as background. All non-$\ttH$ backgrounds 
are considered for both measurements.

\section{Reconstruction of isolated leptons}
\label{sec:isolated_leptons}
The productions cross sections for signal events with six or eight jets in the final 
state are similar. Signal events with six jets contain one high-energetic isolated lepton 
from the leptonic \PW boson decay. On the other hand, no isolated leptons are expected 
in signal events with eight jets. Hence the number of isolated leptons is an important 
observable in the signal selections for both final states. In this study, the isolated electrons 
or muons are considered.

The \texttt{IsolatedLeptonFinder} processor as implemented in
\texttt{MarlinReco}~\cite{marlinreco-homepage} is used to identify leptons in
regions with otherwise little calorimetric activity. The identification of isolated
leptons starts from charged tracks. First isolation criteria and in a second step
identification criteria for electrons and muons based on calorimeter depositions associated to the tracks are applied.
These two steps are discussed in the following.

\begin{figure}
    \centering
    \includegraphics[width=.5\linewidth]{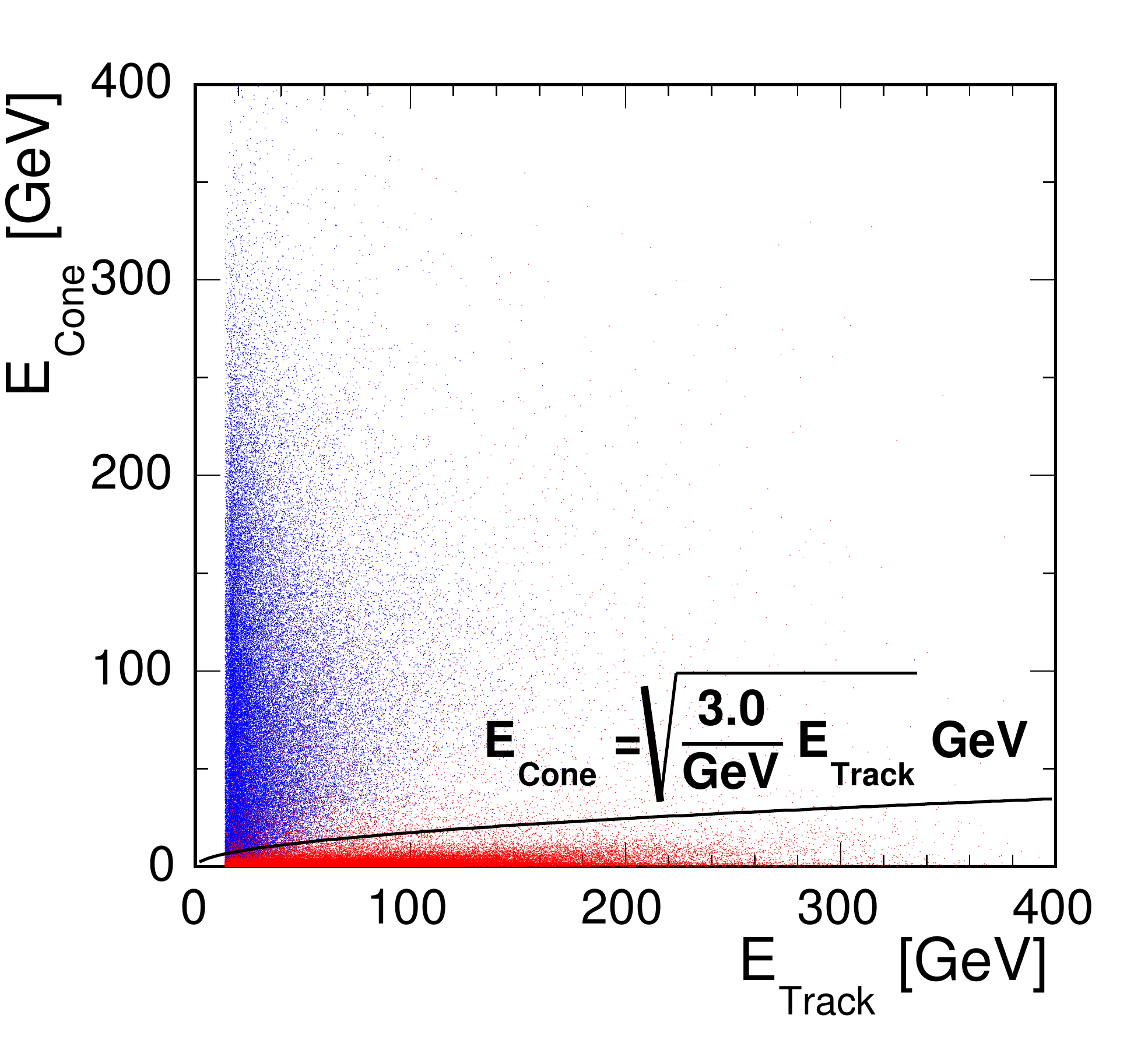}
    \caption{Scatter plot of the energy within a $\cos\theta>0.99$ cone around a track versus the track energy in a sample of \ttH events with one leptonic \PW decay.
    Tracks that are matched to the \PW decay are shown in red, all other tracks are represented by blue points. Tracks below the black curve in this plot are used as isolated lepton candidates.}
    \label{fig:IsoLeptonScatterPlot}
\end{figure}

The isolation criteria
were optimised on a sample of \ttH events with one leptonic \PW decay by a
parameter scan. A set of parameters yielding a high significance for finding
this leptonic \PW decay over selecting other lepton candidates is listed in
Table~\ref{tab:IsolatedLeptonFinderParameters}. This is illustrated in
Figure~\ref{fig:IsoLeptonScatterPlot}, where the energy of particles in a cone
around a track, defined by $\cos\theta>0.99$, is plotted versus the track
energy in a sample of \ttH events with one leptonic \PW decay. The cones
around tracks that are matched to the \PW decay contain generally little
additional energy and the data points are concentrated along the x-axis, while
other tracks are more likely to be found inside a jet and are therefore found
closer to the y- than the x-axis. It is found that a non-linear
parametrisation of the relationship between cone energy and track energy
improves the performance of the isolated lepton identification.

\begin{table}
    \centering
    \begin{subtable}{.49\textwidth}
    \centering
    \caption{Track Parameters}
    \begin{tabular}{l|l}
        \hline \hline
        Parameter & Value \\
        \hline
        2D impact parameter $d_{0}$ & \unit[0.02]{mm}\\
        z-value of POCA $z_{0}$ & \unit[0.1]{mm} \\
        3D impact parameter $r_{0}$ & \unit[0.1]{mm} \\
        \hline\hline
    \end{tabular} 
    \end{subtable}
    \begin{subtable}{.49\textwidth}
    \centering
    \caption{Parameters of the polynomial energy isolation $E_{\text{cone}}^{2} < A E_{\text{lepton}}^{2} + B E_{\text{lepton}} + C$}
    \begin{tabular}{l|l}
        \hline\hline
        cone size & $\cos\theta>0.99$ \\
        A & 0 \\
        B & 3.0 \\
        C & 0 \\
        \hline\hline
    \end{tabular}
    \end{subtable}
    \caption{Chosen set of parameters to steer the \texttt{IsolatedLeptonFinder} processor in \texttt{MarlinReco}. The point of closest approach (POCA) $z_{0}$ is defined as the smallest distance between the track and the primary vertex along the beam direction.}
    \label{tab:IsolatedLeptonFinderParameters}
\end{table}

The electron and muon identification criteria based on their energy deposition
in the ECAL and HCAL were optimised in a separate step. For electrons,
the fraction of ECAL to HCAL energy depositions is required to satisfy \mbox{$0.95
< E_\text{ECAL}/(E_\text{ECAL}+E_\text{HCAL}) < 1$}, and the ratio of ECAL energy deposition
to track energy meets the
requirement \mbox{$0.85 < E_\text{ECAL}/p_\text{track} < 1.15$} (using the pion hypothesis). The
requirements on the energy depositions for the selection of muons are \mbox{$0.03 < E_\text{ECAL}/(E_\text{ECAL}+E_\text{HCAL}) < 0.2$} and \mbox{$0 <
E_\text{ECAL}/p_\text{track} < 0.4$}.

\section{Reconstruction of W, top and Higgs candidates}
\label{sec:reco_w_top_higgs}

As a first step of the event analysis chain, isolated leptons are found
as described in Sec.~\ref{sec:isolated_leptons}. The particle flow objects (PFOs) identified as isolated muons or electrons are excluded from the jet
reconstruction procedure. Only PFOs in the range $20^{\circ} < \theta <
160^{\circ}$ are considered in the following, because the particles originating from the signal processes are 
located in the central part of the detector while the beam-related backgrounds peak in the forward direction (see Sec.~\ref{sec:beam_backgrounds}). The Durham jet clustering algorithm~\cite{Catani1991432} is used in the exclusive mode with six or eight jets.

To form \PW, top and Higgs candidates from the reconstructed jets, the following 
function is minimised for the final state with eight jets:
\begin{equation}
\chi^{2}_{\textnormal{8 jets}} = \frac{(M_{12}-M_{\PW})^{2}}{\sigma_{\PW}^{2}} + \frac{(M_{123}-M_{\PQt})^{2}}{\sigma_{\PQt}^{2}} + \frac{(M_{45}-M_{\PW})^{2}}{\sigma_{\PW}^{2}} + \frac{(M_{456}-M_{\PQt})^{2}}{\sigma_{\PQt}^{2}} + \frac{(M_{78}-M_{\smH})^{2}}{\sigma_{\smH}^{2}},
\label{sid:benchmarking:eq:tth_eight_jet_pairing}
\end{equation}
where $M_{12}$ and $M_{45}$ are the invariant masses of the jet pairs used to
reconstructed the \PW candidates, $M_{123}$ and $M_{456}$ are the invariant
masses of the three jets used to reconstruct the top candidates and $M_{78}$ is
the invariant mass of the jet pair used to reconstruct the Higgs candidate.
$M_{\PW}$, $M_{t}$ and $M_{\smH}$ are the nominal \PW, top and Higgs masses.
The resolutions $\sigma_{\PW}$, $\sigma_{\text{t}}$ and $\sigma_{\smH}$ were
obtained from reconstructed jet combinations matched to \PW, top and Higgs particles at 
generator level. The corresponding function minimised for the six jet final state
is given by:
\begin{equation}
\chi^{2}_{\textnormal{6 jets}} = \frac{(M_{12}-M_{\PW})^{2}}{\sigma_{\PW}^{2}} + \frac{(M_{123}-M_{\PQt})^{2}}{\sigma_{\PQt}^{2}} + \frac{(M_{45}-M_{\smH})^{2}}{\sigma_{\smH}^{2}}.
\label{sid:benchmarking:eq:tth_six_jet_pairing}
\end{equation}

\section{Beauty identification}

\begin{figure}[h]
    \centering
    \includegraphics[width=.49\textwidth]{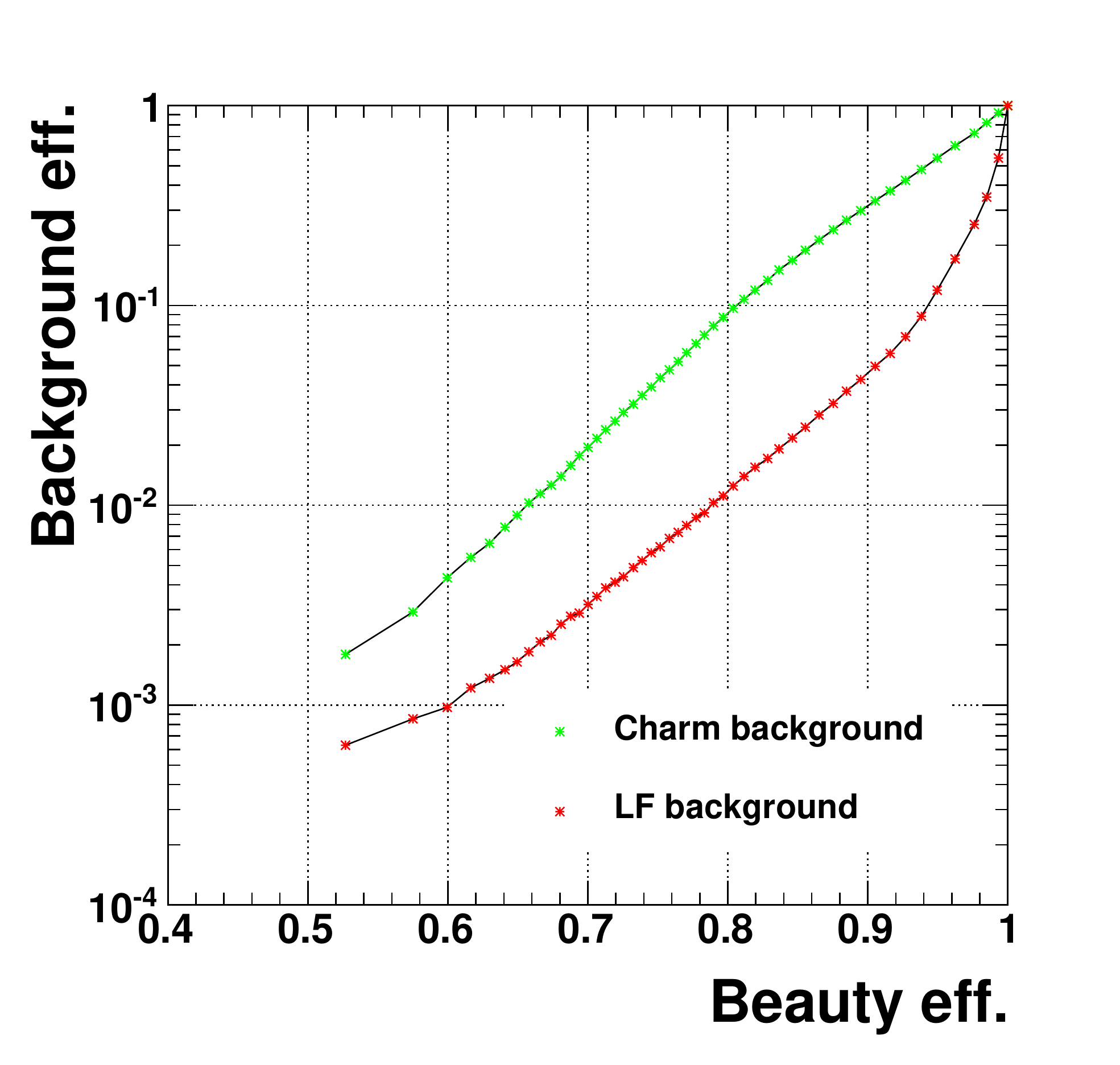}
    \includegraphics[width=.49\textwidth]{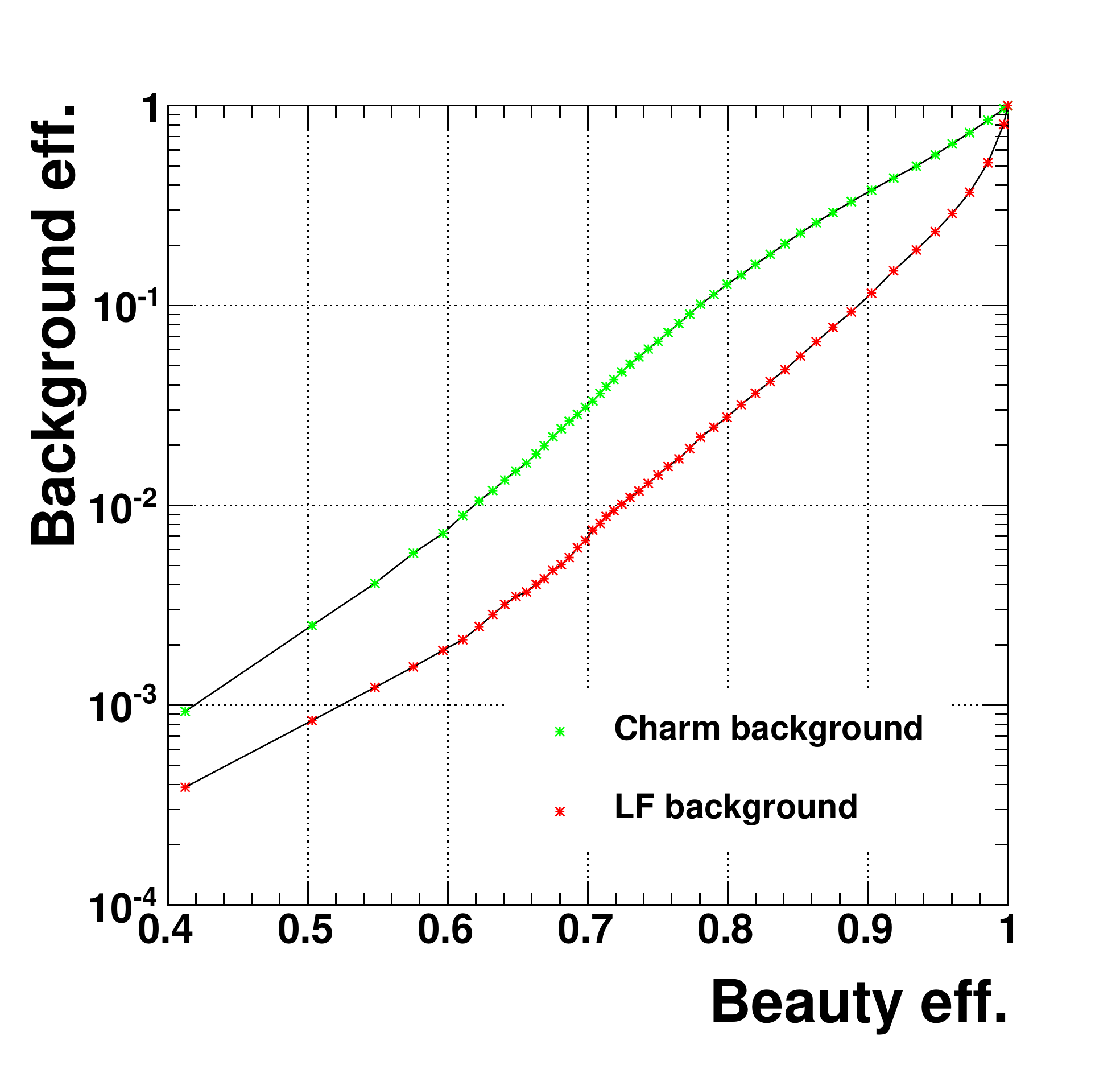}
    \caption{Mis-identification efficiency of light flavour (LF) quark jets (red points) and charm jets (green points) as beauty jets versus beauty identification efficiency in di-jets at \mbox{$\sqrt{s} = \unit[91]{GeV}$}. The performance is shown without (left) and with (right) background from \gghadrons events and incoherent pairs.}
    \label{fig:flavour_tagging_performance}
\end{figure}

A b-tag value is obtained for each jet reconstructed as described in the section above. To illustrate the flavour tagging performance of the \texttt{sidloi3} detector, the mis-identification efficiency of light quark (u, d and s) jets or charm jets, respectively, as beauty jets versus the beauty identification efficiency is shown in Figure~\ref{fig:flavour_tagging_performance}. The mis-identification rates are calculated using di-jet events at a centre-of-mass energy of $\unit[91.2]{GeV}$. For this figure, no cut on the polar angles of the PFOs used as input to the jet reconstruction is applied. A sizeable degradation of the flavour tagging performance due to the impact of beam-induced backgrounds is observed. This effect is smaller for the mis-identification of charm jets than for light quark jets.

The training of the flavour tagging for the measurement of \ttH production reported in this document is based on events with six quarks of the same flavour produced in electron-positron annihilation. For the training, 60000 charm- and beauty-jets, and 180000 light quark jets are used. These samples were chosen since the jets have similar kinematic properties as those in \ttH signal events.

\section{Event selection}

Events were selected using Boosted Decision Trees (BDTs) as implemented in TMVA~\cite{Hoecker2007}.
The BDTs were trained separately for the
eight and six jet final states. The following input variables were used:
\begin{itemize}
\item the four highest b-tag values. The third and fourth highest b-tag value are especially suited to reject $\toppair$ and most of the $\toppair Z$ events which contain only two b-jets;
\item the event thrust. Since the top quarks in $\toppair$ events are produced back to back, the thrust variable has larger values in $\toppair$ events compared to $\ttH$, $\toppair Z$ or $\toppair \bpair$ events;
\item a distance value from the Durham algorithm. The distance parameter between $i$ and \mbox{$j = (i - 1)$} jets is defined as:
\begin{equation}
Y_{ij} = \frac{\min(E_{m}^{2},E_{n}^{2})(1 - \cos\theta_{mn})}{s},
\end{equation}
where $m$ and $n$ are chosen to minimise the distance of the two jets which are merged. If more jets are reconstructed than coloured final state partons are present in an event, the distance parameter tends to small values. For the six jet final state $Y_{65}$ is used while $Y_{87}$ is used for the eight jet final state;
\item the number of reconstructed PFOs in the range $20^{\circ} < \theta < 160^{\circ}$;
\item the number of identified isolated electrons and muons;
\item the missing transverse momentum, $p_\mathrm{T}^{\textnormal{miss}}$, calculated from the reconstructed jets. Due to the leptonic \PW boson decay, finite values of $p_\mathrm{T}^{\textnormal{miss}}$ are reconstructed for six jet signal events while $p_\mathrm{T}^{\textnormal{miss}}$ tends towards zero for eight jet signal events;
\item the total visible energy defined as the scalar sum of all jet energies;
\item the masses $M_{12}$, $M_{123}$ and $M_{45}$ as defined in Section~\ref{sec:reco_w_top_higgs}.
\end{itemize}
For the eight jet final state two additional variables are included:
\begin{itemize}
\item{$M_{456}$ and $M_{78}$ as defined in Section~\ref{sec:reco_w_top_higgs}.}
\end{itemize}

The distributions of all input variables are shown in 
Appendix~\ref{sec:appendix_6jets} for the six jet final state and in 
Appendix~\ref{sec:appendix_8jets} for the eight jet final state.

\begin{figure}[h]
\includegraphics[width=0.45\textwidth]{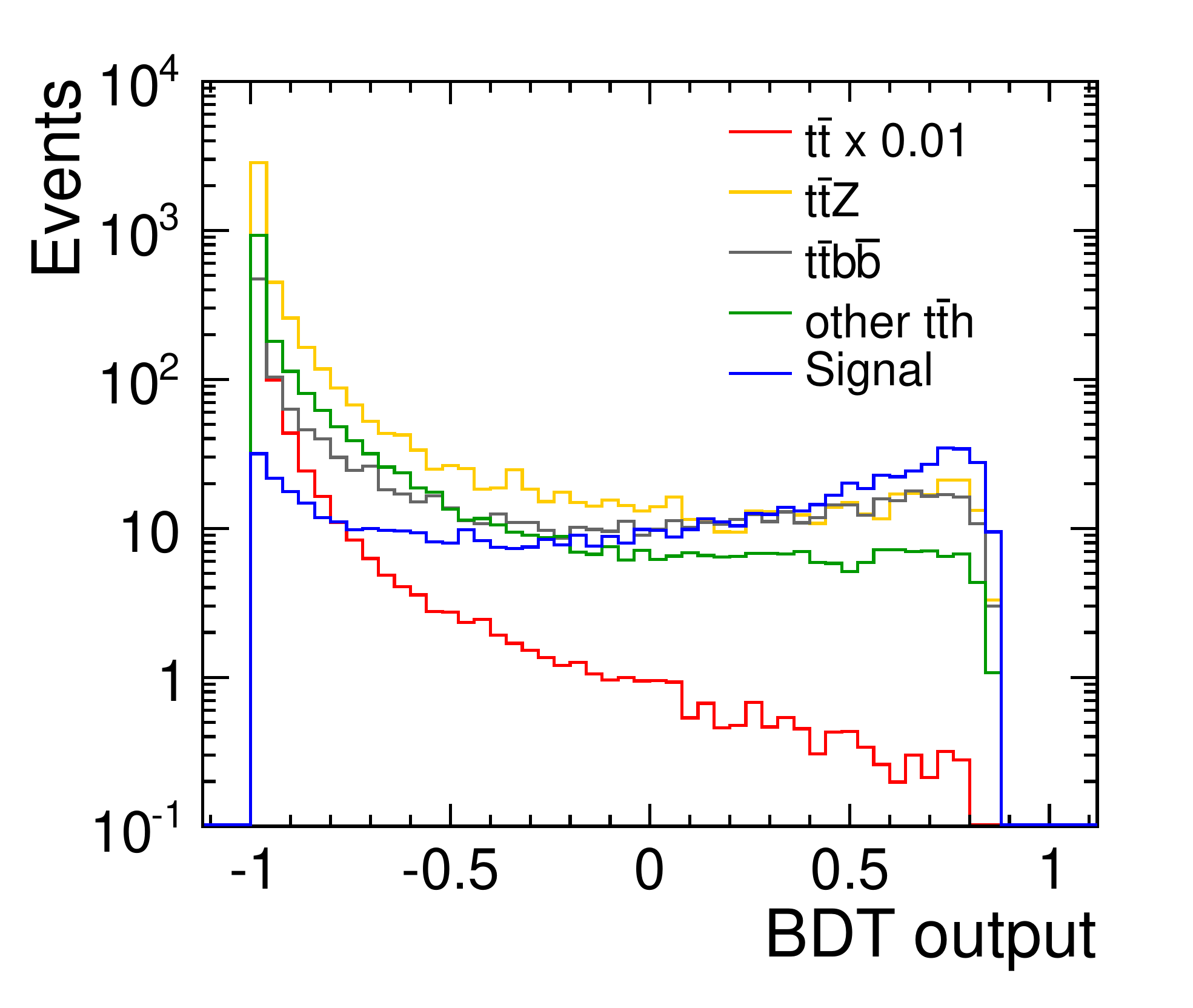} \hspace{0.5cm}
\includegraphics[width=0.45\textwidth]{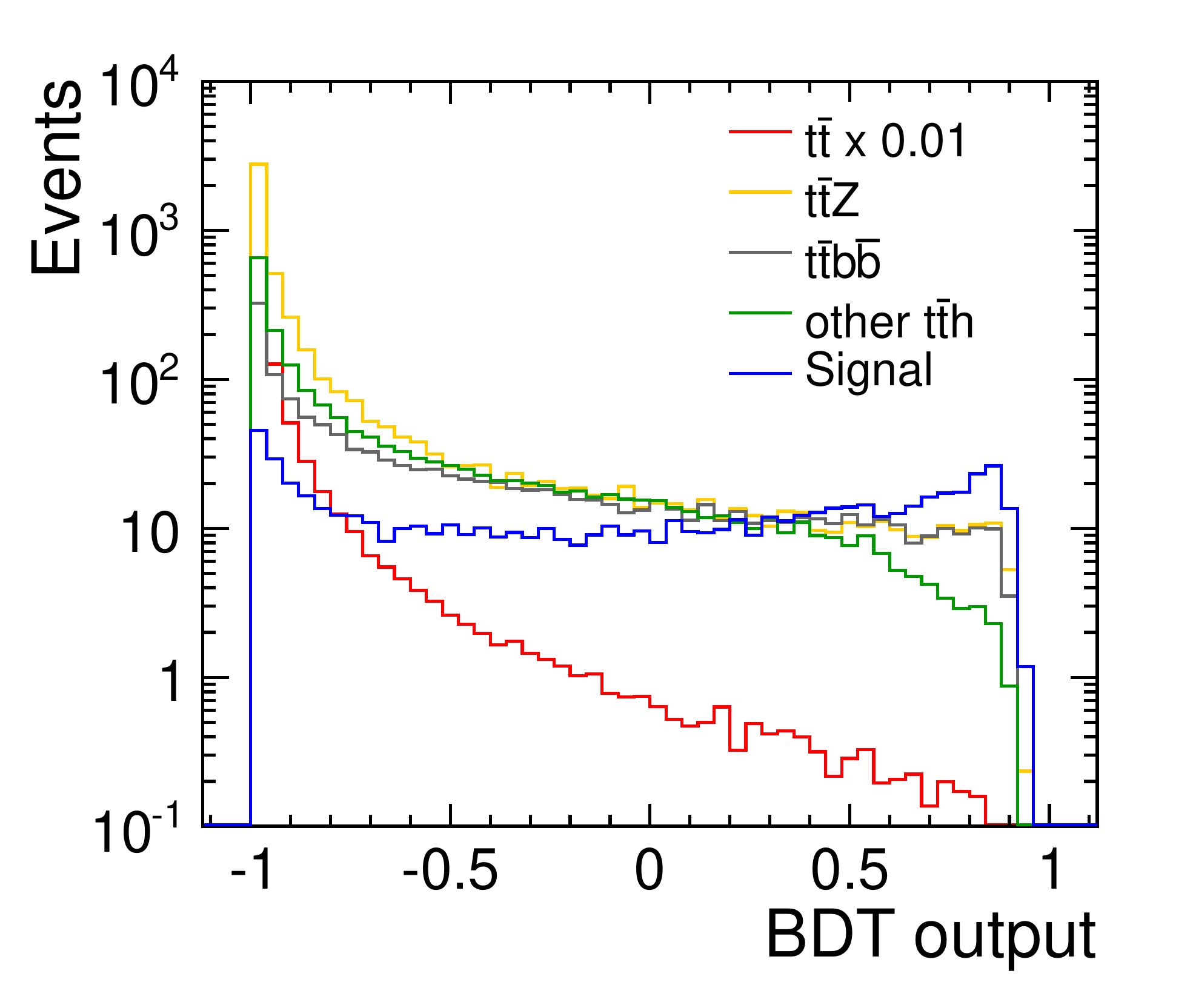}
\caption{\label{sid:benchmarking:fig:bt_outputs_tth} Output distributions of the BDTs for the 
eight (left) and six (right) jet final states. The signals are shown in blue while the 
backgrounds are shown in different colours. The distribution for \toppair was scaled by a 
factor of $0.01$.}
\end{figure}

\begin{figure}[h]
\includegraphics[width=0.45\textwidth]{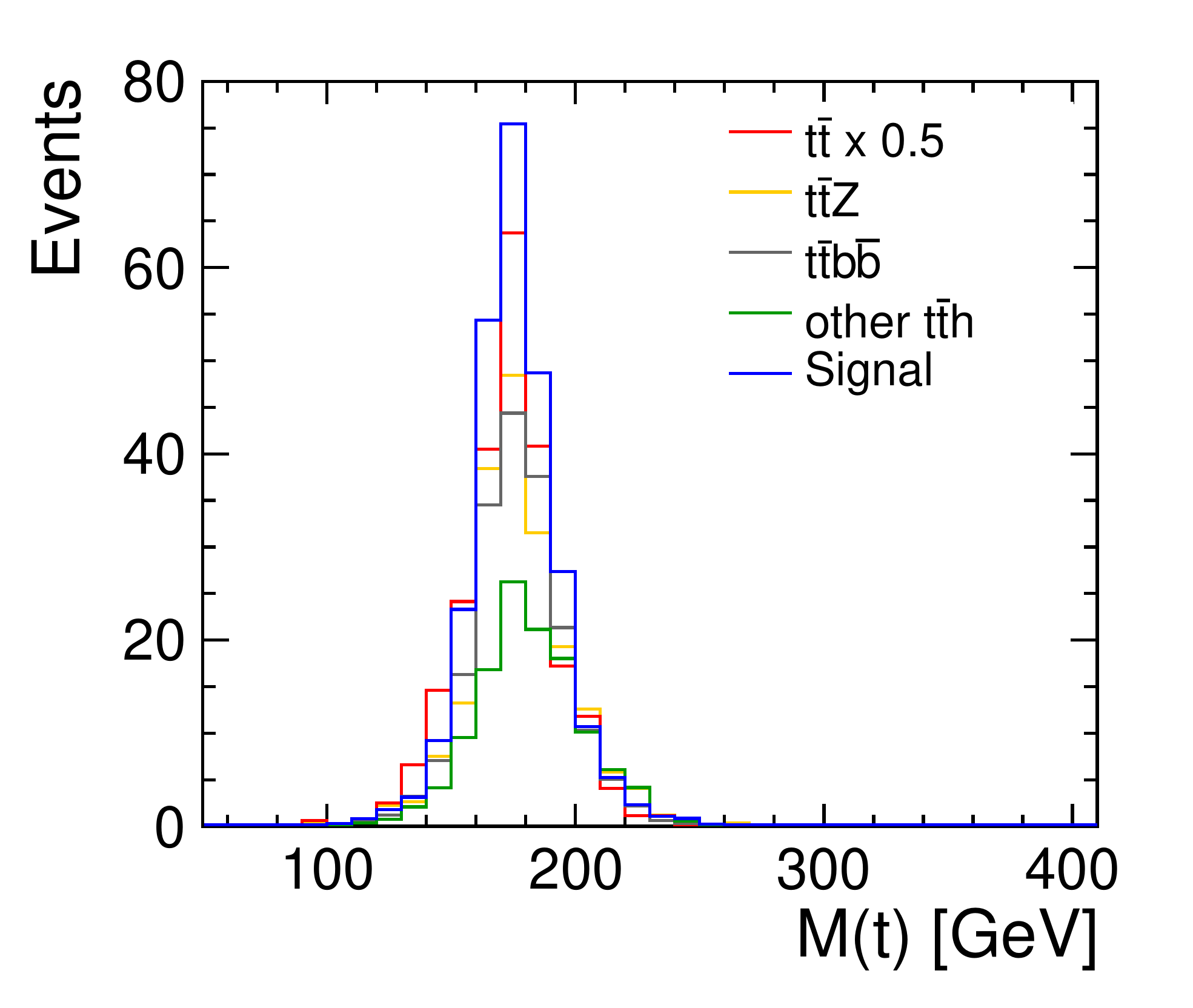} \hspace{0.5cm}
\includegraphics[width=0.45\textwidth]{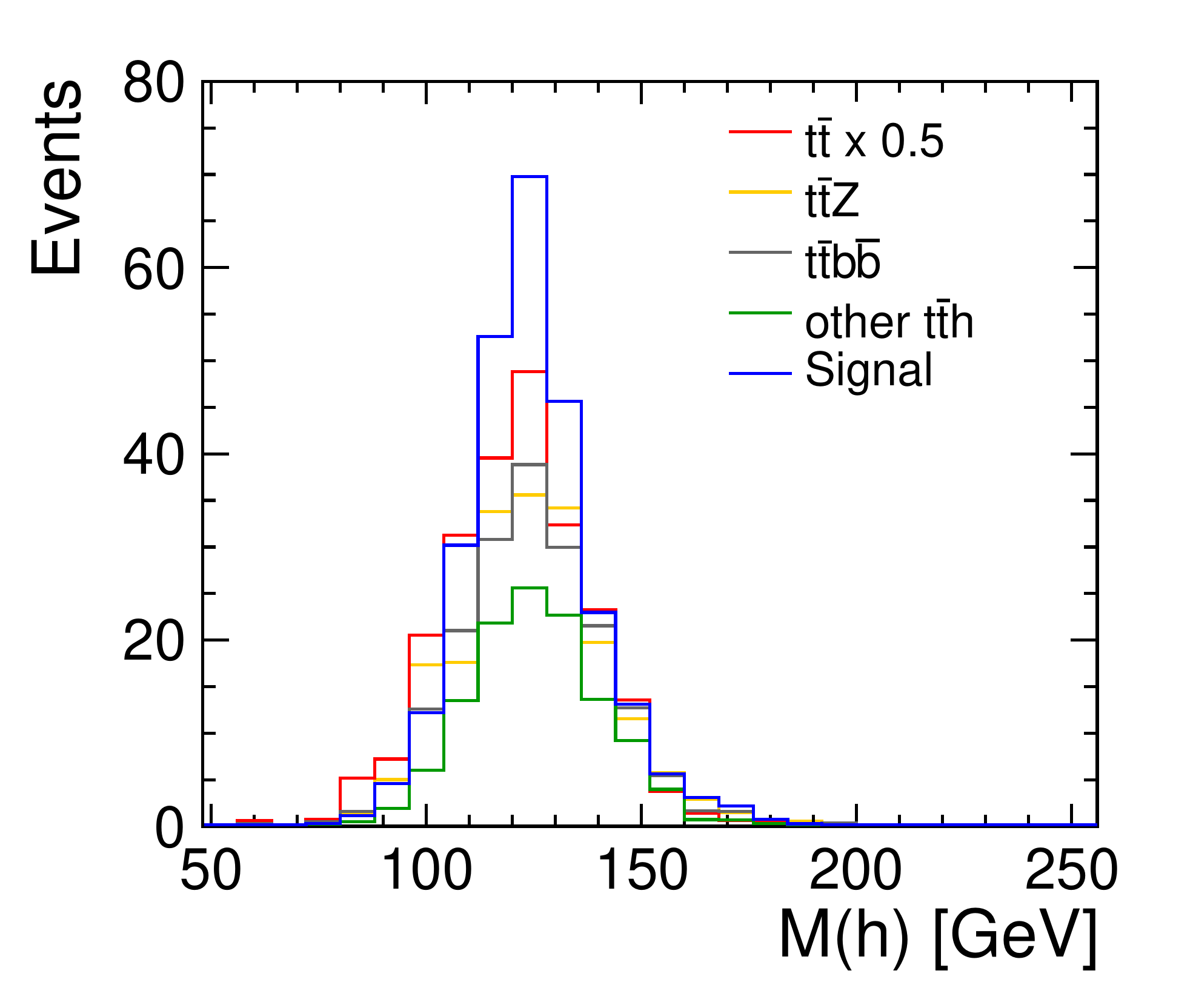}
\caption{\label{sid:benchmarking:fig:reconstructed_masses_tth} Reconstructed top (left) 
and Higgs (right) masses for the selected (BDT output value $>$ 0.1978) six jet events. The signal is shown in blue while the 
backgrounds are shown in different colours. The distribution for \toppair was scaled by a 
factor of $0.5$.}
\end{figure}

The output values of the BDTs for the signals and for the different backgrounds are shown in
Figure~\ref{sid:benchmarking:fig:bt_outputs_tth} for both final states.
To select events, cuts on the BDT output values are applied. The cuts were
optimised by maximising the signal significance given by:
$\frac{S}{\sqrt{S+B}}$, where $S$ is the number of signal events and $B$ is the
number of background events. As an example, the reconstructed top and Higgs masses in 
six jet events after the cut on the BDT output are shown in Figure~\ref{sid:benchmarking:fig:reconstructed_masses_tth}. 
The selection efficiencies for signal events are 42\% and 54\% for the six and 
eight jet final states, respectively. In Table~\ref{tab:tth_selected_events} the numbers 
of events passing the cuts on the BDT output values are shown separately for all
investigated final states.

\begin{table}
\caption{\label{tab:tth_selected_events} Number of selected events for the different 
final states assuming an integrated luminosity of 1~\abinv. The values obtained for the six 
and eight jet final state selections are shown separately.}
\begin{center}
\begin{tabular}{l R[.]{3}{1} R[.]{3}{1}} \toprule
Final state & \multicolumn{1}{c}{BDT trained to select 6 jets} & \multicolumn{1}{c}{BDT trained to select 8 jets} \\ \midrule
\ttH, $\smH \to \bpair$ (6 jets) & 264.9 & 87.2 \\
\ttH, $\smH \to \bpair$ (8 jets) & 72.6 & 356.2 \\
\ttH, $\smH$ not $\bpair$ (6 jets) & 11.7 & 5.1 \\
\ttH, $\smH$ not $\bpair$ (8 jets) & 4.3 & 21.6 \\
\ttH (4 jets) & 32.8 & 2.1 \\
$\toppair Z$ & 188.4 & 253.6 \\
$\toppair g^{*} \to \toppair\bpair$ & 185.0 & 243.6 \\
$\toppair$ & 459.3 & 687.0 \\ \bottomrule
\end{tabular}
\end{center}
\end{table}

\section{Results}
The cross section can be directly obtained from the number of
background-subtracted signal events after the selection. The 
uncertainty of the cross section measurement was estimated using the number 
of selected signal and background events. Assuming an
integrated luminosity of 1~\abinv split equally between the \mbox{$P(\Pem) = -80\%$}, \mbox{$P(\Pep) = +20\%$} 
and \mbox{$P(\Pem) = +80\%$}, \mbox{$P(\Pep) = -20\%$} beam polarisaiton configurations, the cross section can be measured with a
statistical accuracy of $11.5\%$ using the eight jet final state and with a
statistical accuracy of $13.2\%$ for the six jet final state.

As a cross check, the analysis was repeated preselecting events with one isolated lepton for the 
six jet final state and events without isolated leptons for the eight jet final state. In 
Appendix~\ref{sec:appendix_selected_events_preselection} the numbers of selected events 
are shown for this approach. 
The differences in precision on the top Yukawa coupling compared to the nominal analysis are negligible.

To extract the top Yukawa coupling from the measured cross sections, signal Monte Carlo samples 
with different values of the top Yukawa coupling were generated. The dependence of 
the cross section on the value of the coupling was fitted using a quadratic function. The 
following relation was found: $\frac{\Delta y_{\PQt}}{y_{\PQt}} = 0.52 \cdot \frac{\Delta \sigma}{\sigma}$~\cite{lc-rep-2013-004}. 
The factor between the cross section uncertainty and the coupling uncertainty differs 
from $0.5$ due to the contribution from Higgsstrahlung to the $\ttH$ production 
cross section. The uncertainties of the measured cross sections translate to
precisions on the top Yukawa coupling of $6.0\%$ and $6.9\%$ from the eight and
six jet final states, respectively. If both measurements are combined, the top
Yukawa coupling can be extracted with a statistical accuracy of $4.5\%$. Good agreement 
with a similar study performed using the ILD detector concept~\cite{lc-rep-2013-004} is observed.

For 1~\abinv of data with only $P(\Pem) = -80\%$, $P(\Pep) = +20\%$ polarisation, this 
number improves to $4.0\%$.

The precision for the six jet final state could be improved further if 
$\tau$-leptons were included in the reconstruction. Additional improvements of the analysis 
procedure like kinematic fitting will be investigated in the future.

The uncertainty on BR($\smH \to \bpair$) is neglected in the calculation of the top Yukawa coupling
from the $\ttH$ production cross section, because it is expected that this quantity 
can be measured with a precision of better than $1\%$ using $\epluseminus \to \PGn\PAGn\smH$ 
events~\cite{Barklow:2003hz, lc-rep-2013-005}.

Systematic uncertainties were not investigated in detail so far. However, it is expected that the relevant sources 
of systematic uncertainty like the beauty identification, the jet energy scale or the knowledge of 
the luminosity spectrum will result in errors that are small compared to the statistical precision of the measurements. The 
understanding of the detector acceptances can be checked using processes like $\toppair$ or $\toppair Z$ 
where the cross sections can be predicted precisely.

\section{Summary}
The physics potential for a measurement of the top Yukawa coupling at \unit[1]{TeV} using the SiD detector is investigated. The study is based on a full detector simulation. Beam-induced backgrounds are considered in the analysis. The combination of results obtained for two different final states leads to a statistical uncertainty on the top Yukawa coupling of $4.5\%$ for an integrated luminosity of 0.5~\abinv with the \mbox{$P(\Pem) = -80\%$, $P(\Pep) = +20\%$} beam polarisation configuration and 0.5~\abinv with \mbox{$P(\Pem) = +80\%$, $P(\Pep) = -20\%$} polarisation.

\section{Acknowledgements}
The authors would like to thank Tim Barklow, Mikael Berggren and Akiya Miyamoto for generating the Monte Carlo samples. We gratefully acknowlege the help from Christian Grefe and Stephane Poss with the production on the Grid. Finally, we wish to thank Tony Prince and Tomohiko Tanabe for comparisons of the results to the ILD analysis.

\newpage

\appendix

\section{Control plots for the six jet final state}
\label{sec:appendix_6jets}

\begin{figure}[H]
\includegraphics[width=0.45\textwidth]{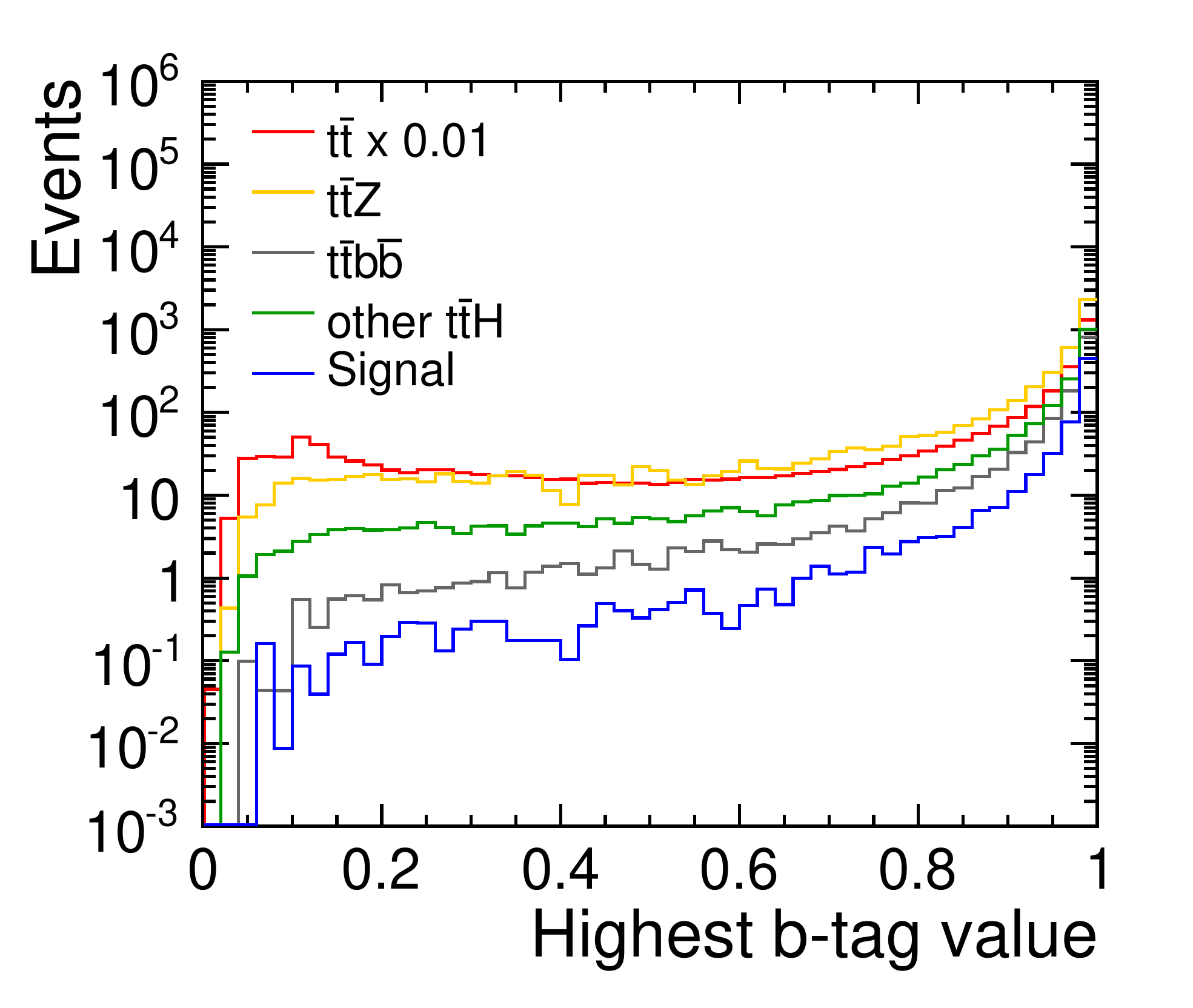} \hspace{0.5cm}
\includegraphics[width=0.45\textwidth]{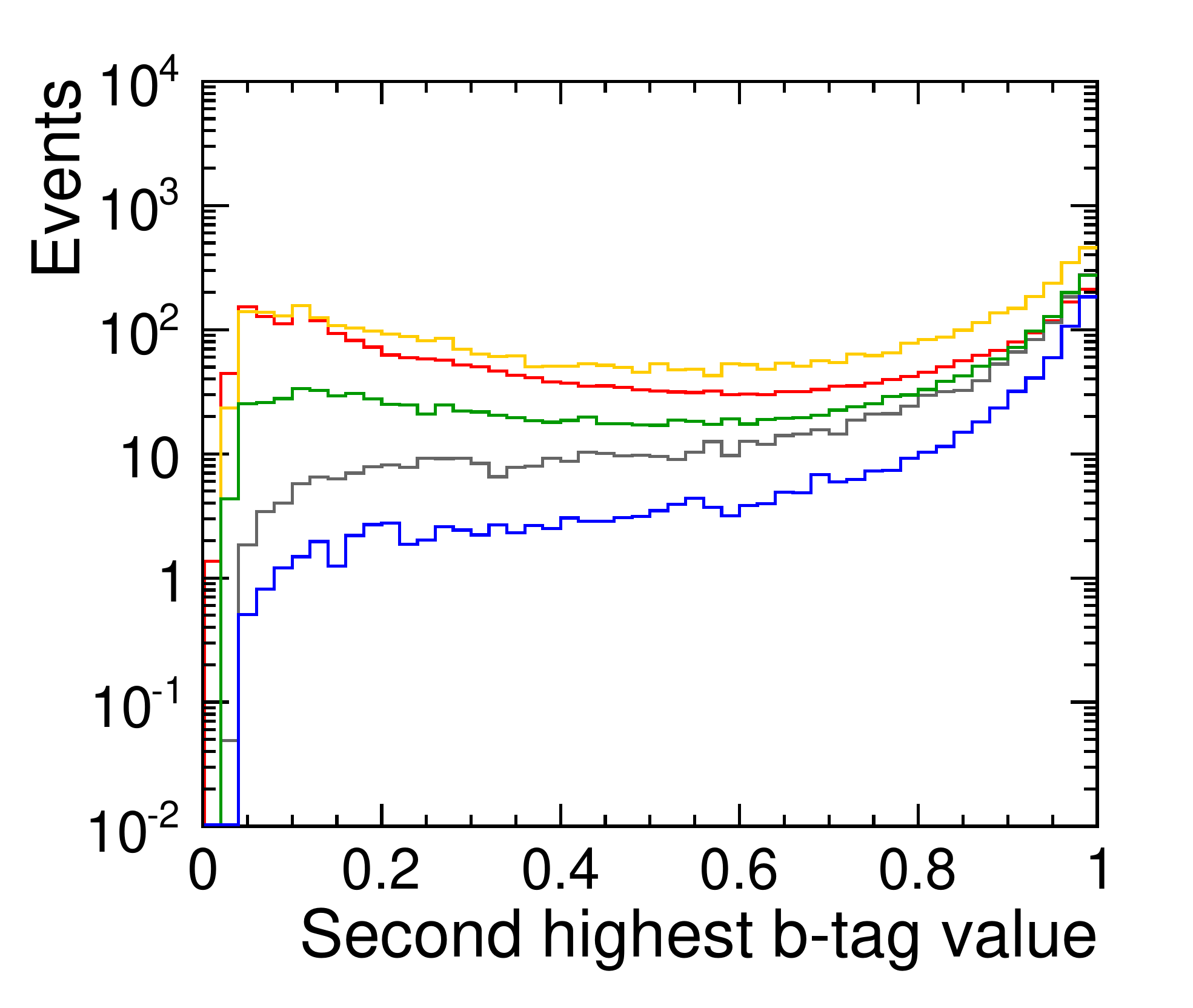} \\
\includegraphics[width=0.45\textwidth]{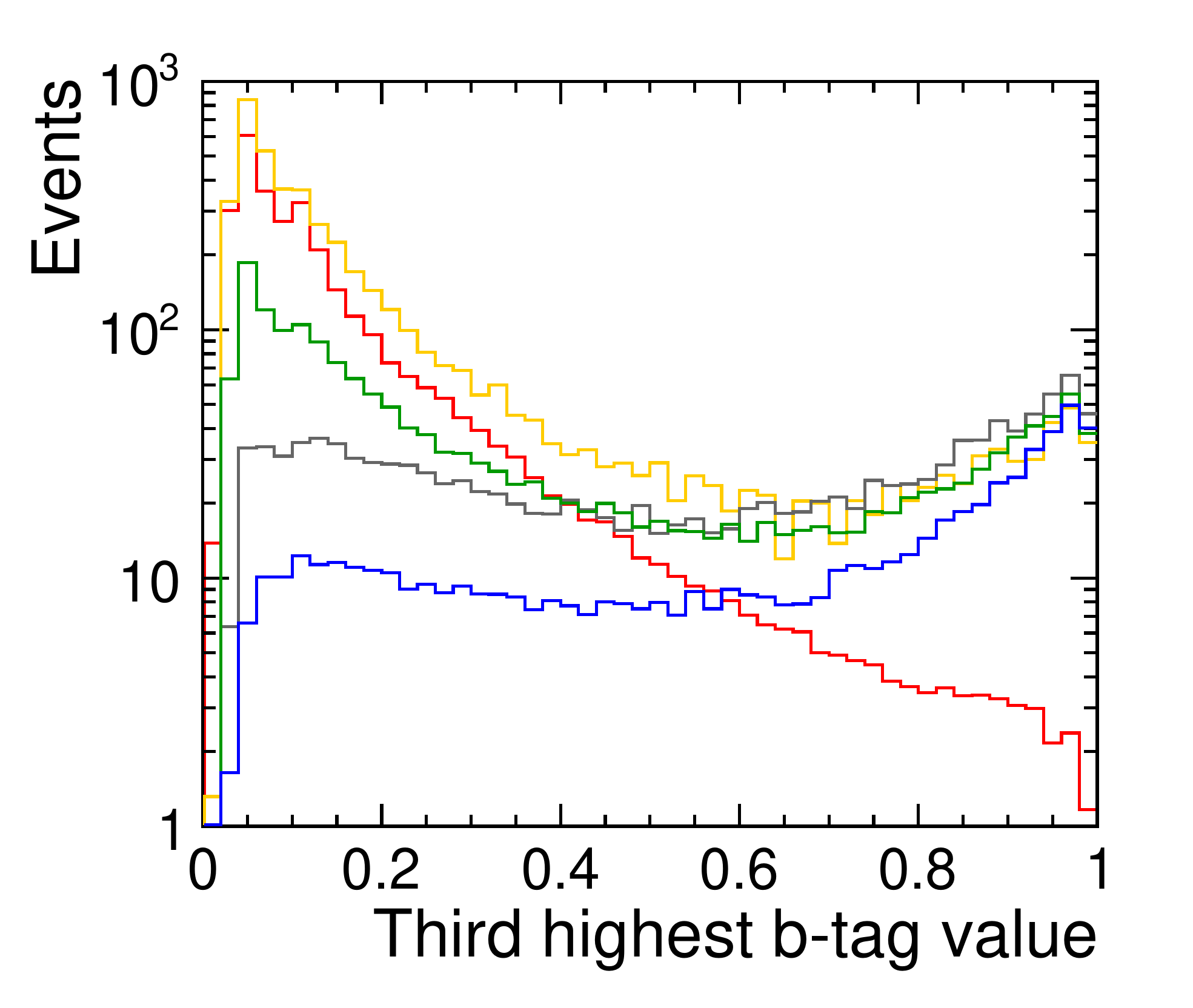} \hspace{0.5cm}
\includegraphics[width=0.45\textwidth]{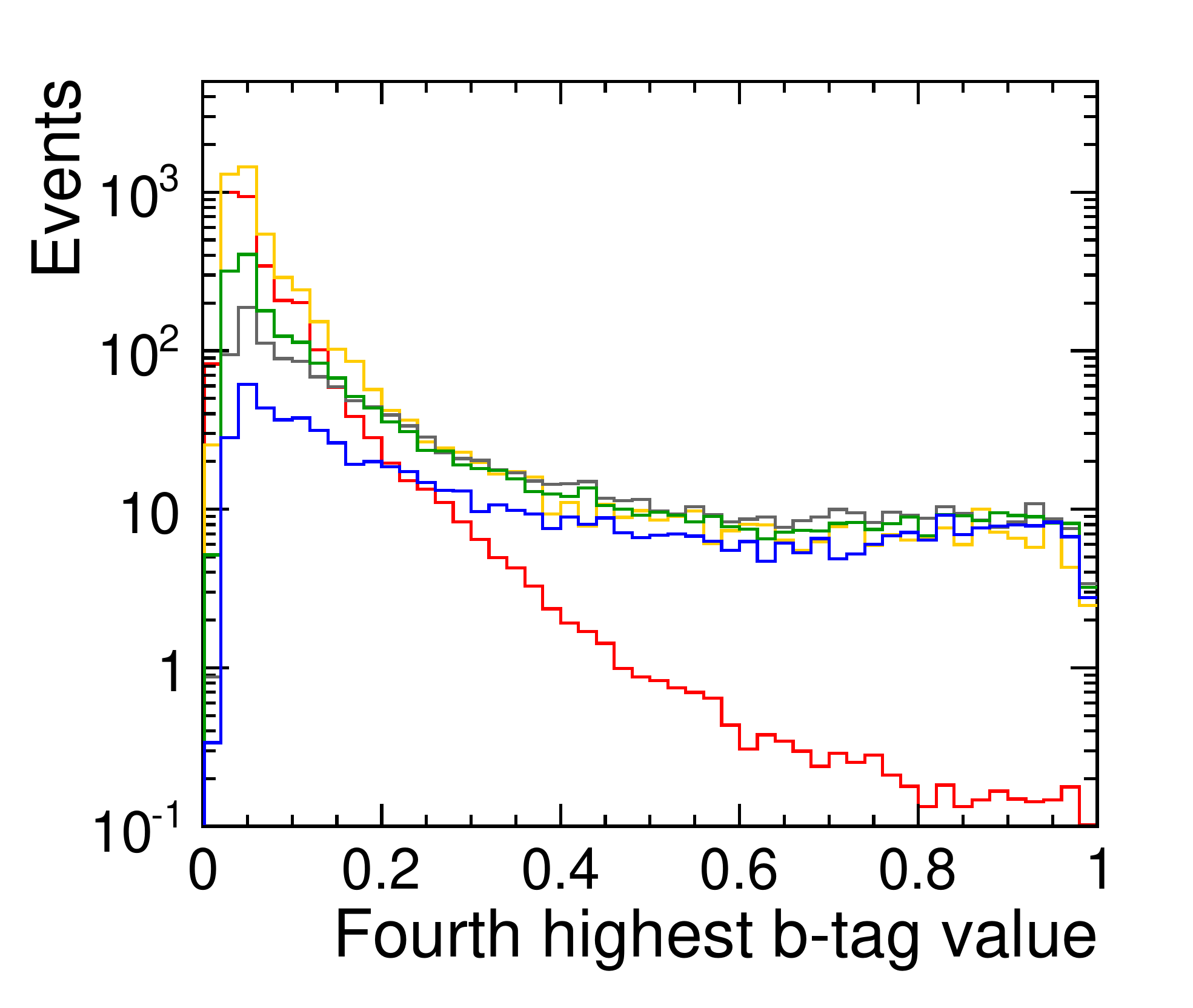} \\
\caption{\label{sid:benchmarking:fig:bt_outputs_tth_sixJets_btag} Distributions of several discriminating
variables used in the event selection for the six jet final state. The signals are shown in blue while the 
backgrounds are shown in different colours. The distribution for \toppair was scaled by a factor of $0.01$.}
\end{figure}

\begin{figure}[H]
\includegraphics[width=0.45\textwidth]{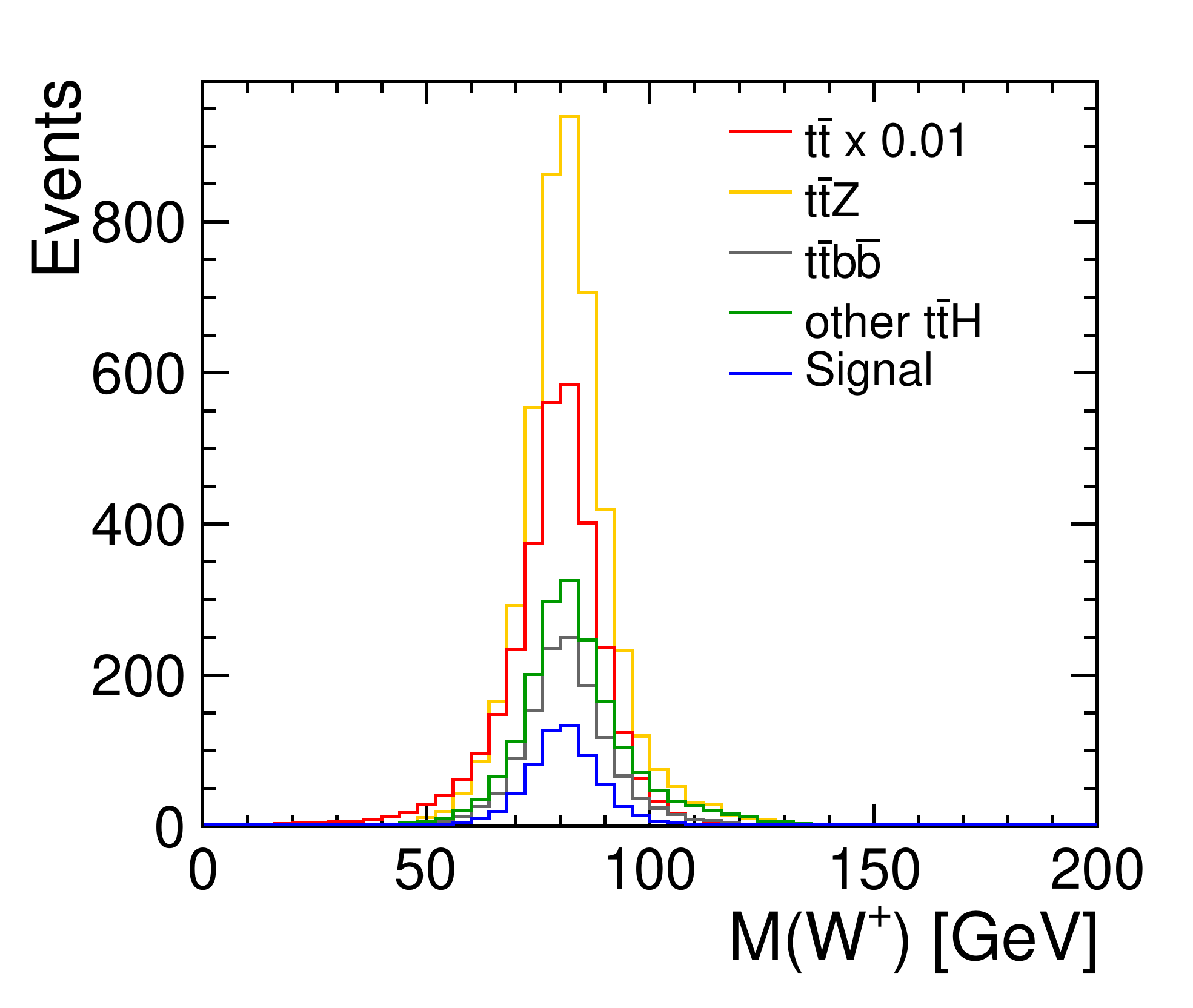} \hspace{0.5cm}
\includegraphics[width=0.45\textwidth]{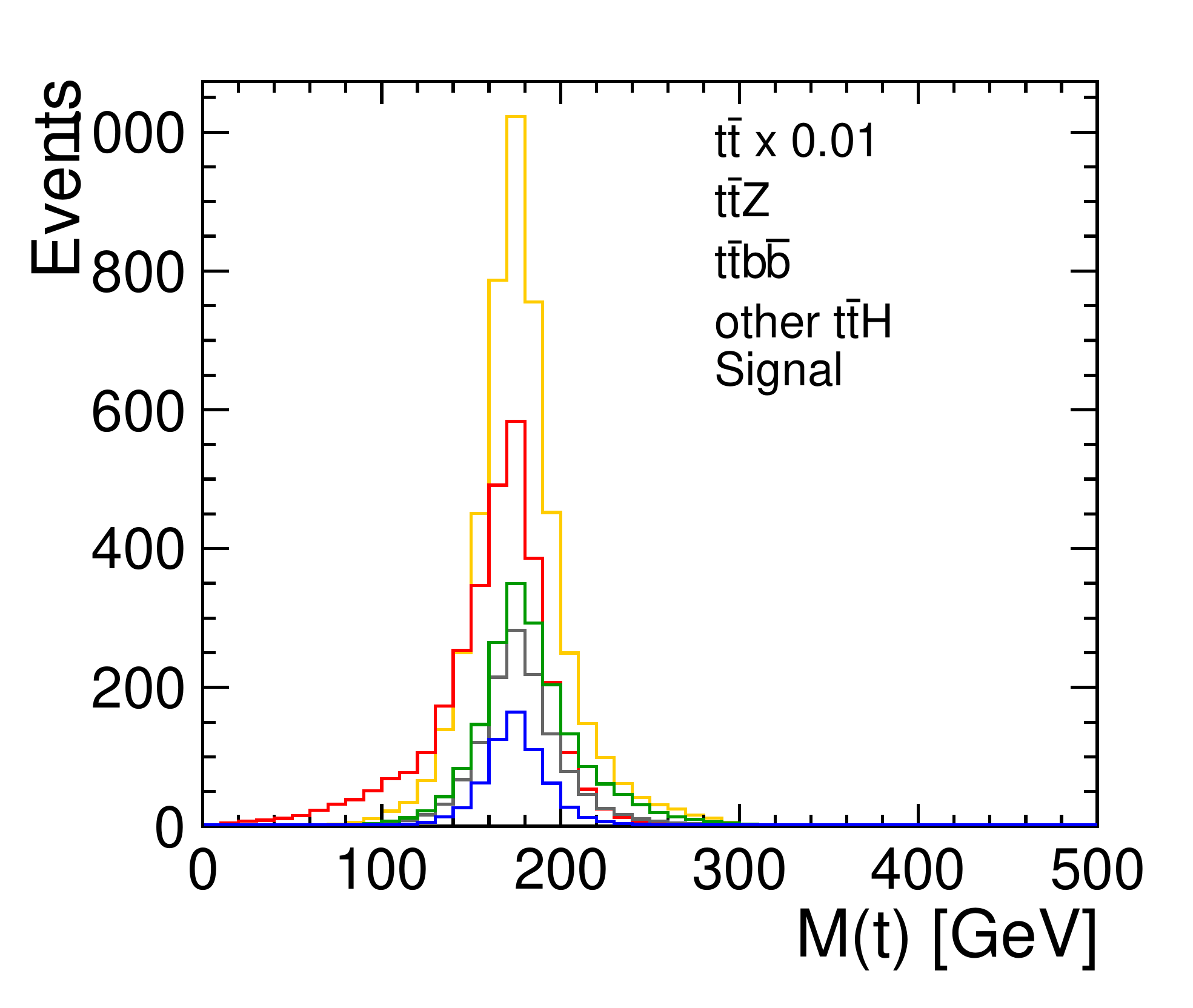} \\
\includegraphics[width=0.45\textwidth]{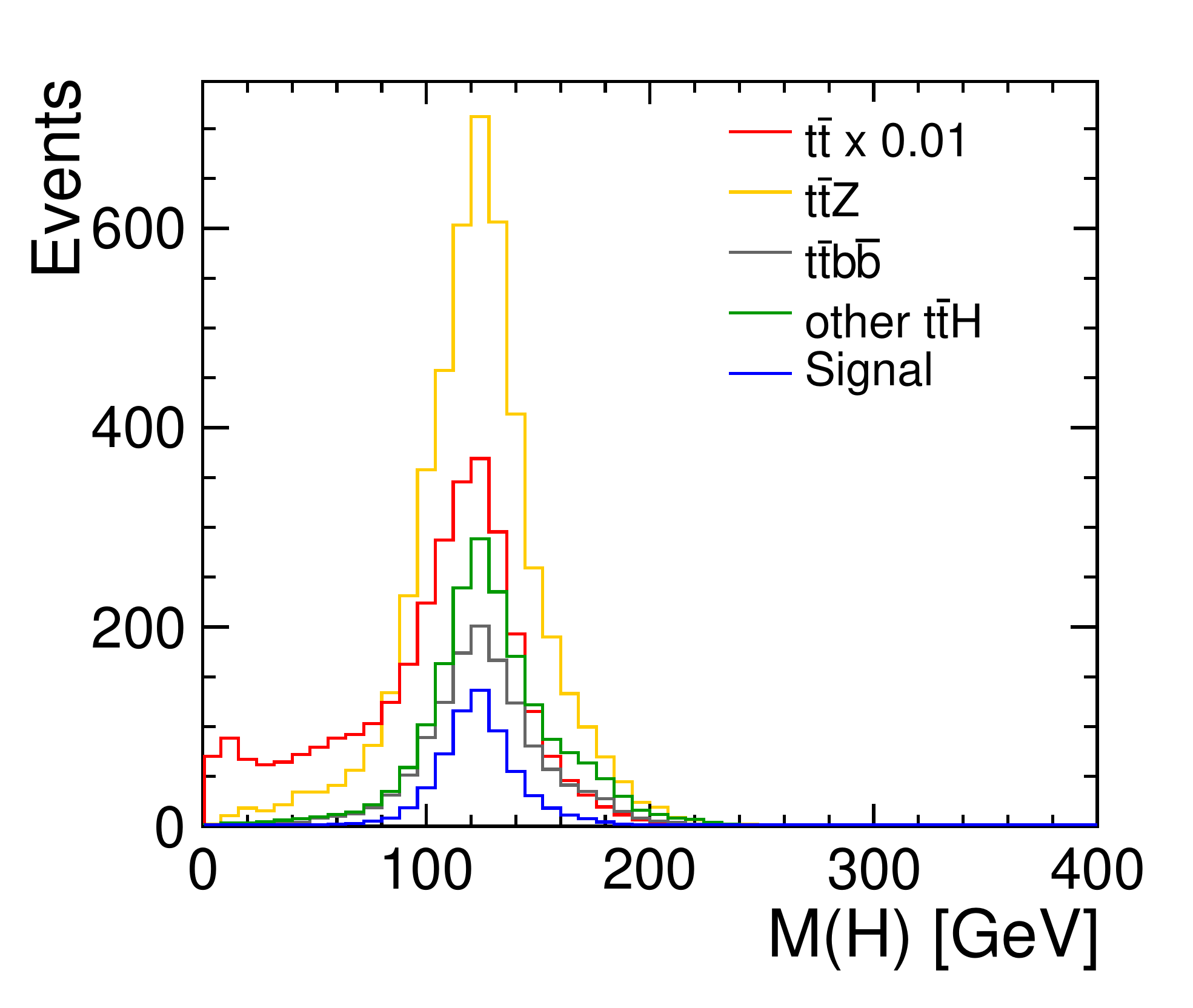} \hspace{0.5cm}
\includegraphics[width=0.45\textwidth]{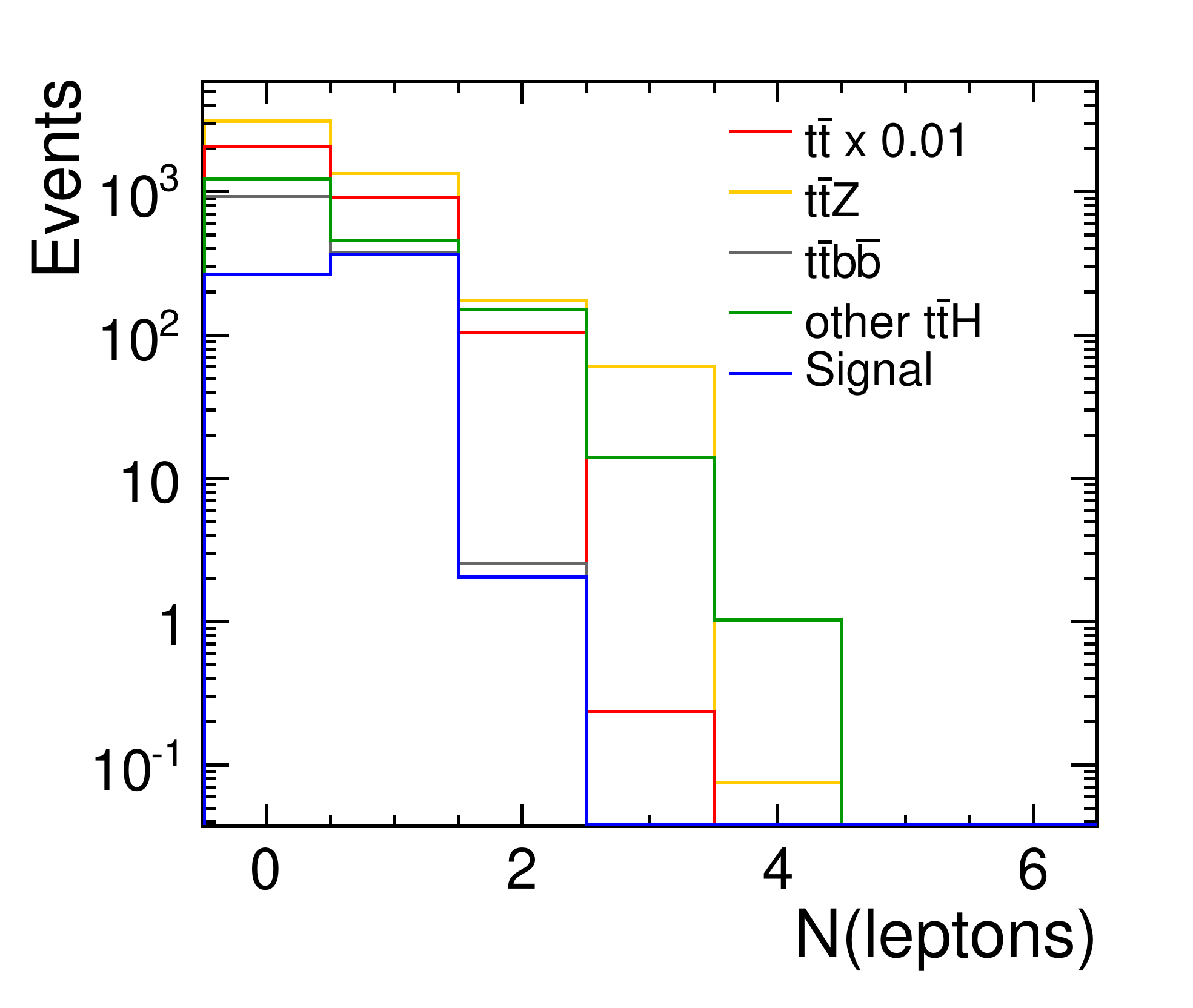} \\
\includegraphics[width=0.45\textwidth]{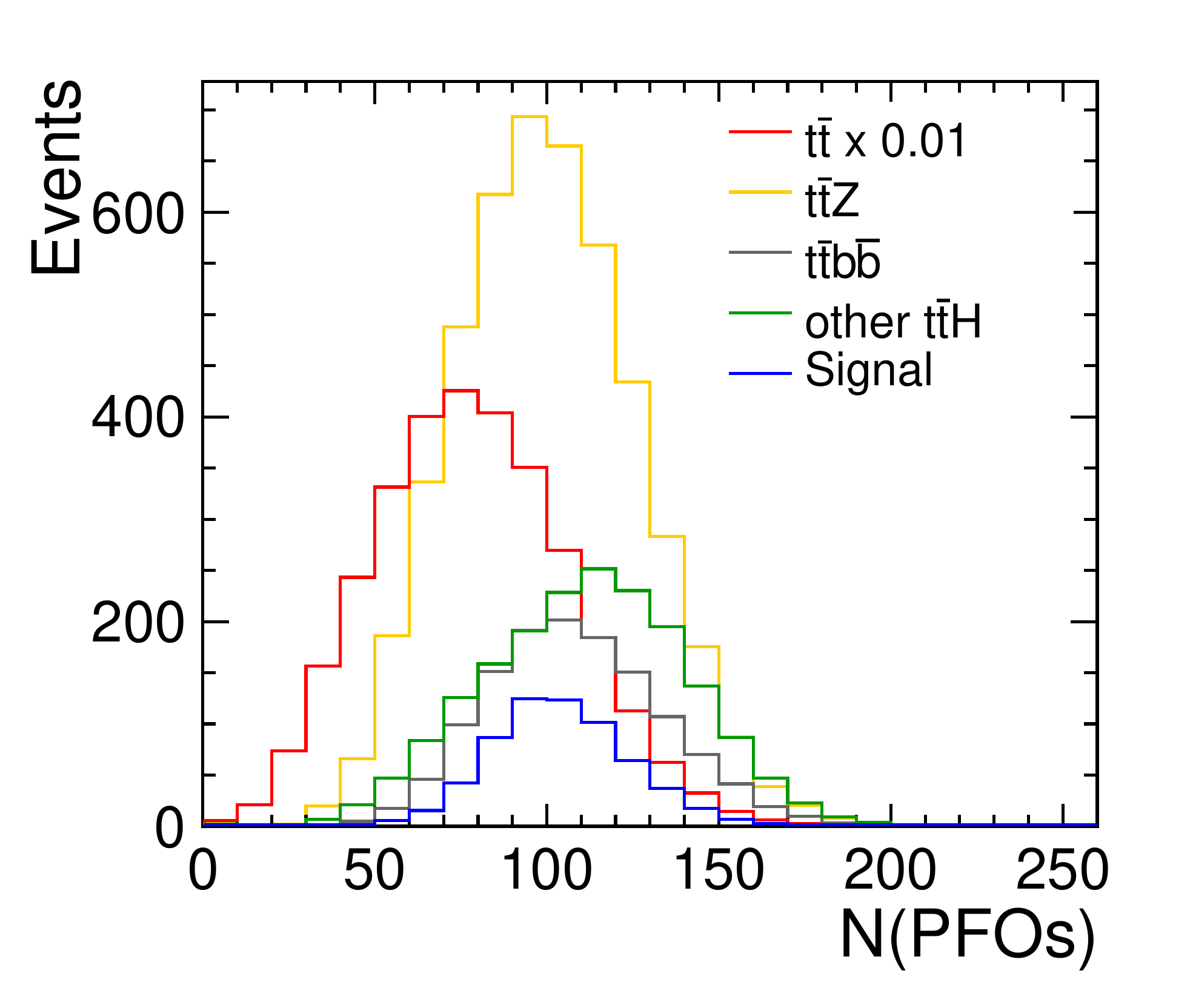} \hspace{0.5cm}
\includegraphics[width=0.45\textwidth]{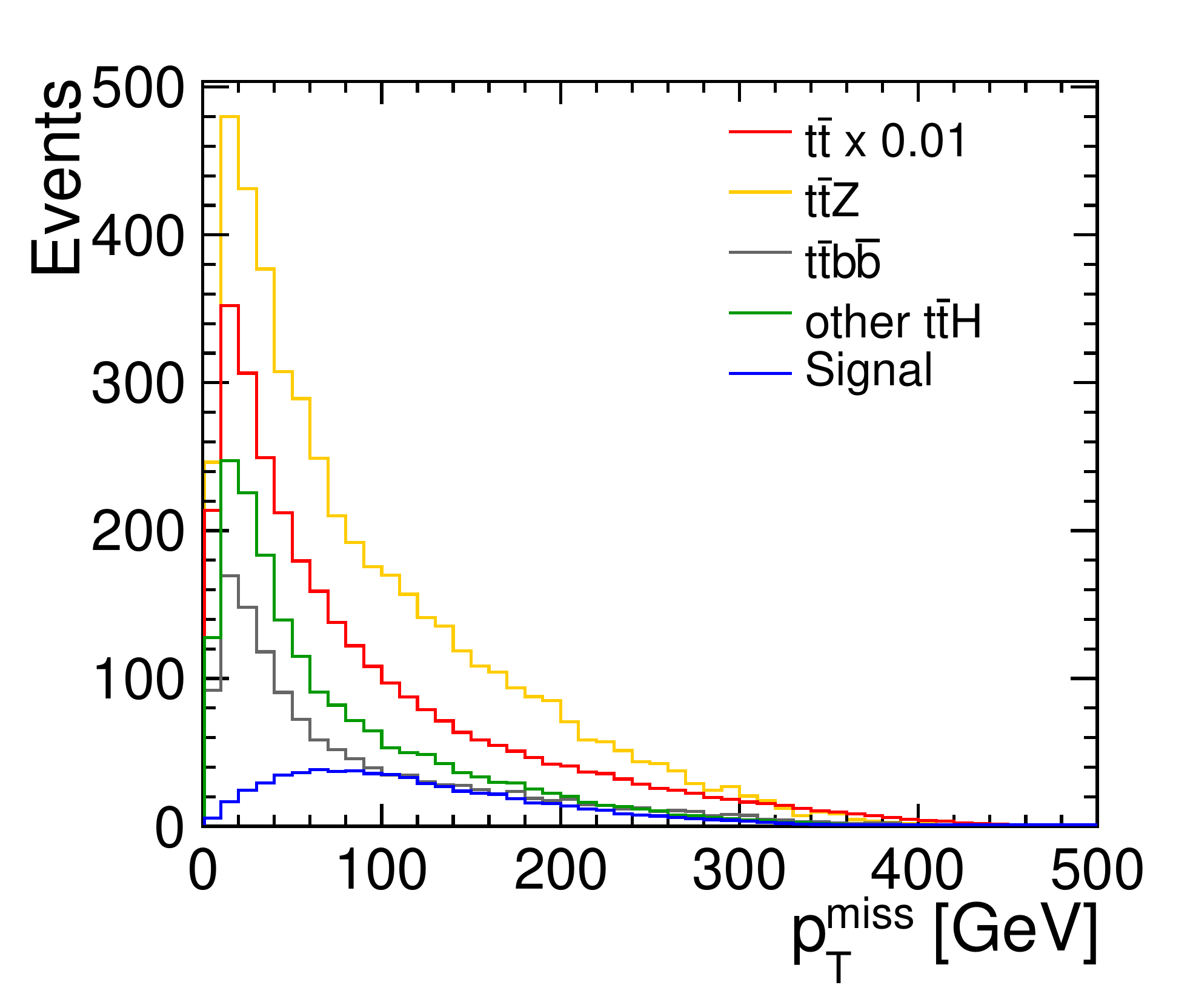} \\
\caption{\label{sid:benchmarking:fig:bt_outputs_tth_sixJets_mass} Distributions of several discriminating  
variables used in the event selection for the six jet final state. The signals are shown in blue while the 
backgrounds are shown in different colours. The distribution for \toppair was scaled by a factor of $0.01$.}
\end{figure}

\begin{figure}[H]
\includegraphics[width=0.45\textwidth]{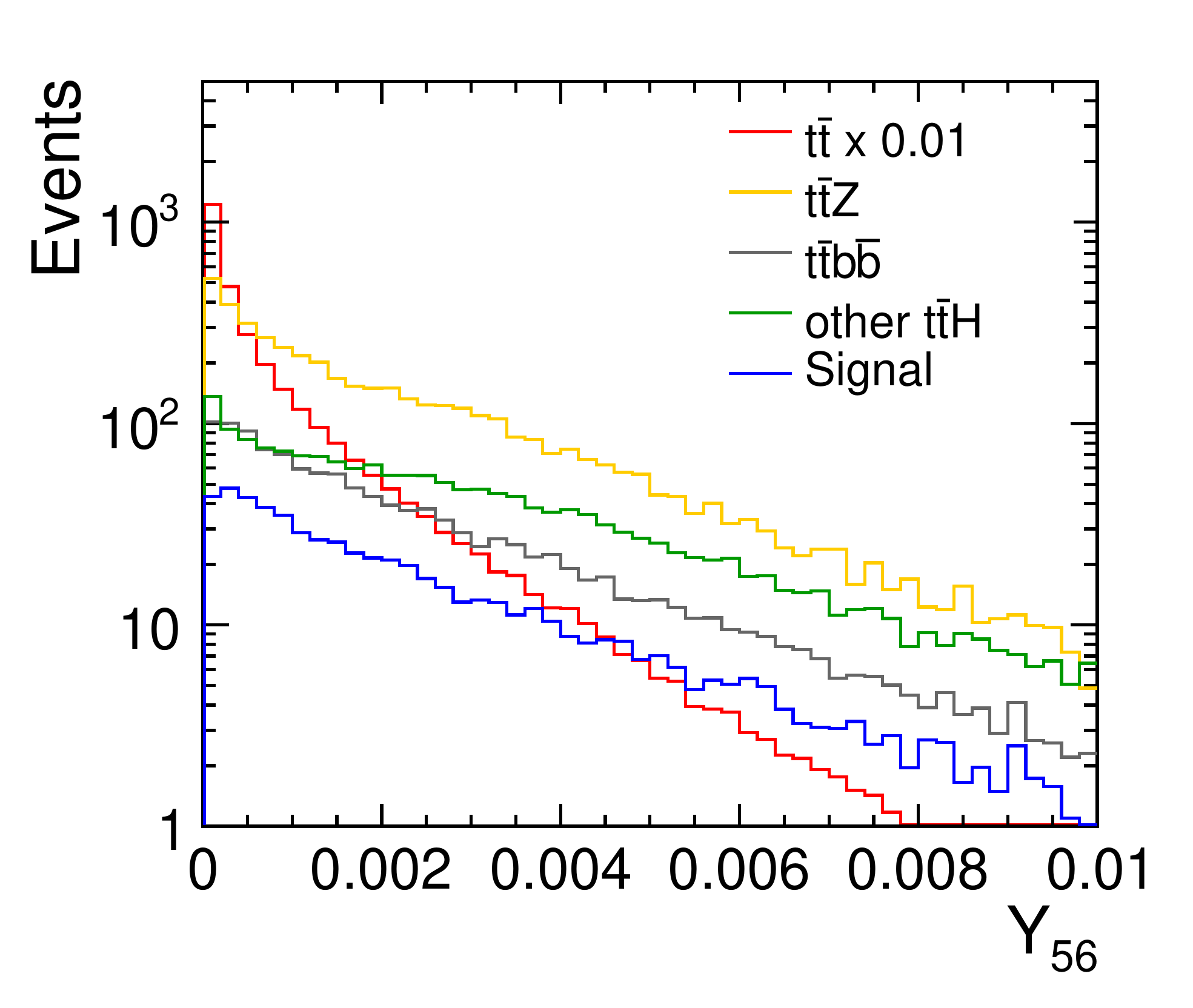} \hspace{0.5cm}
\includegraphics[width=0.45\textwidth]{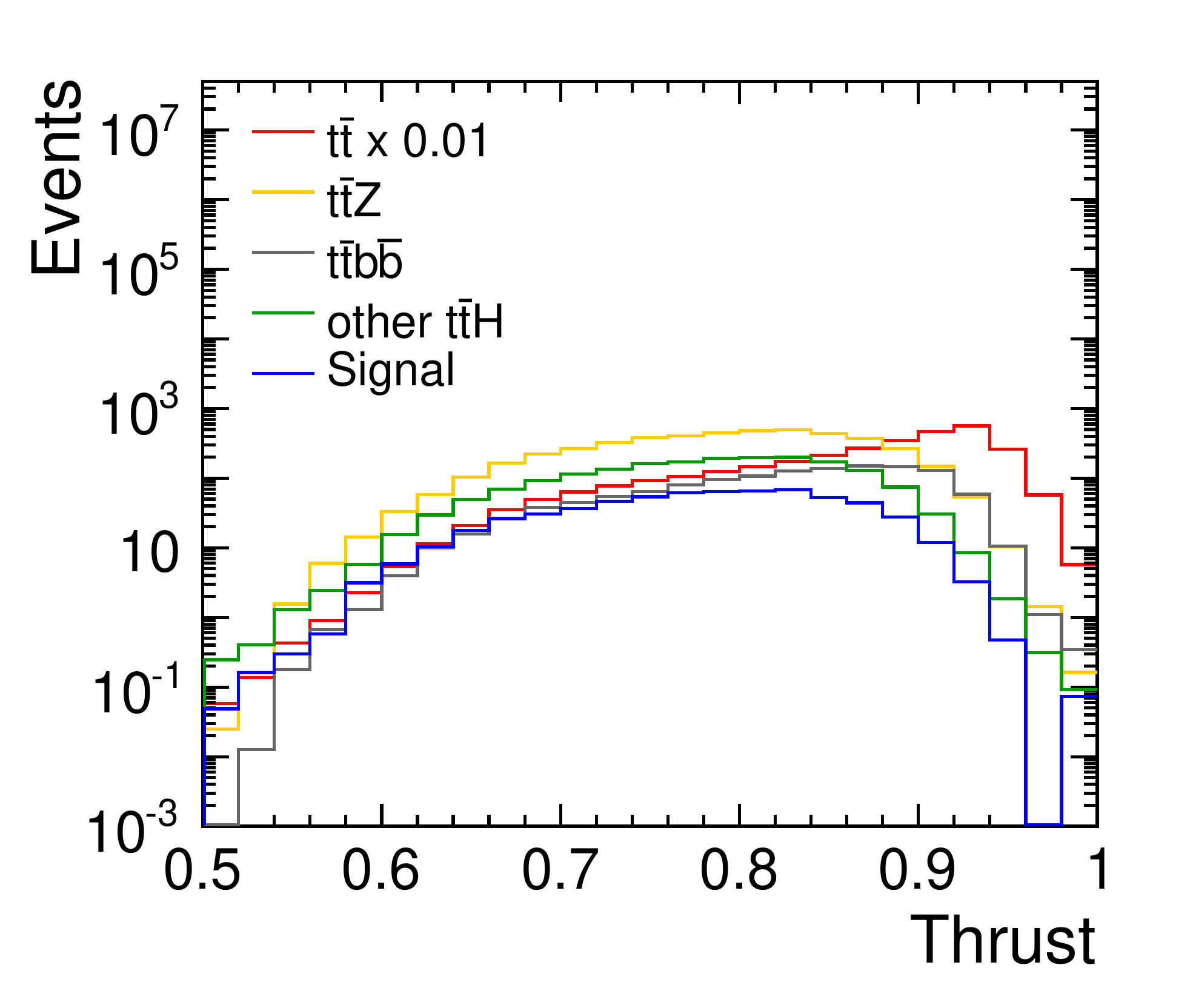} \\
\includegraphics[width=0.45\textwidth]{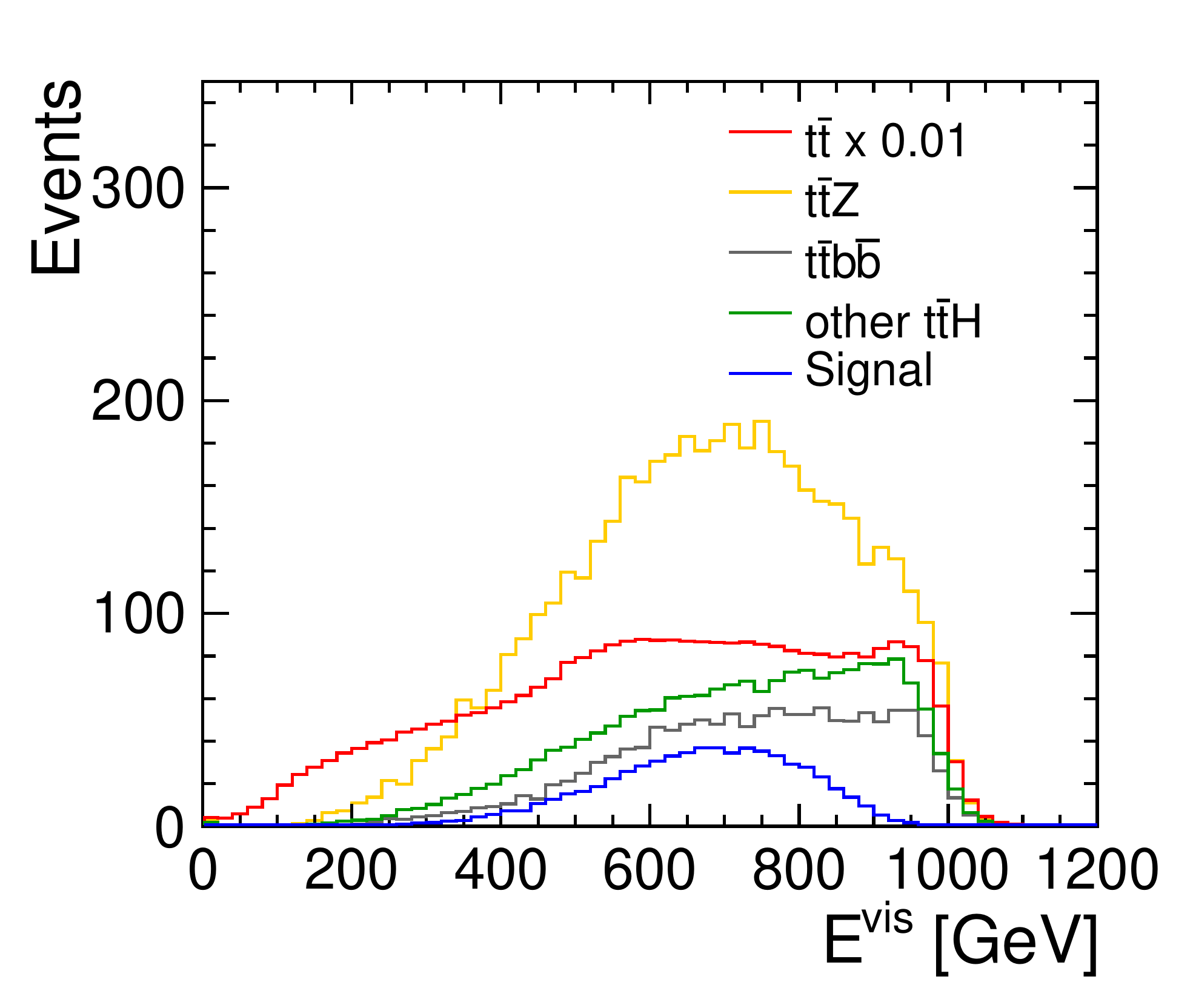} \hspace{0.5cm}
\caption{\label{sid:benchmarking:fig:bt_outputs_tth_sixJets_eventShape} Distributions of several discriminating  
variables used in the event selection for the six jet final state. The signals are shown in blue while the 
backgrounds are shown in different colours. The distribution for \toppair was scaled by a factor of $0.01$.}
\end{figure}

\newpage

\section{Control plots for the eight jet final state}
\label{sec:appendix_8jets}

\begin{figure}[H]
\includegraphics[width=0.45\textwidth]{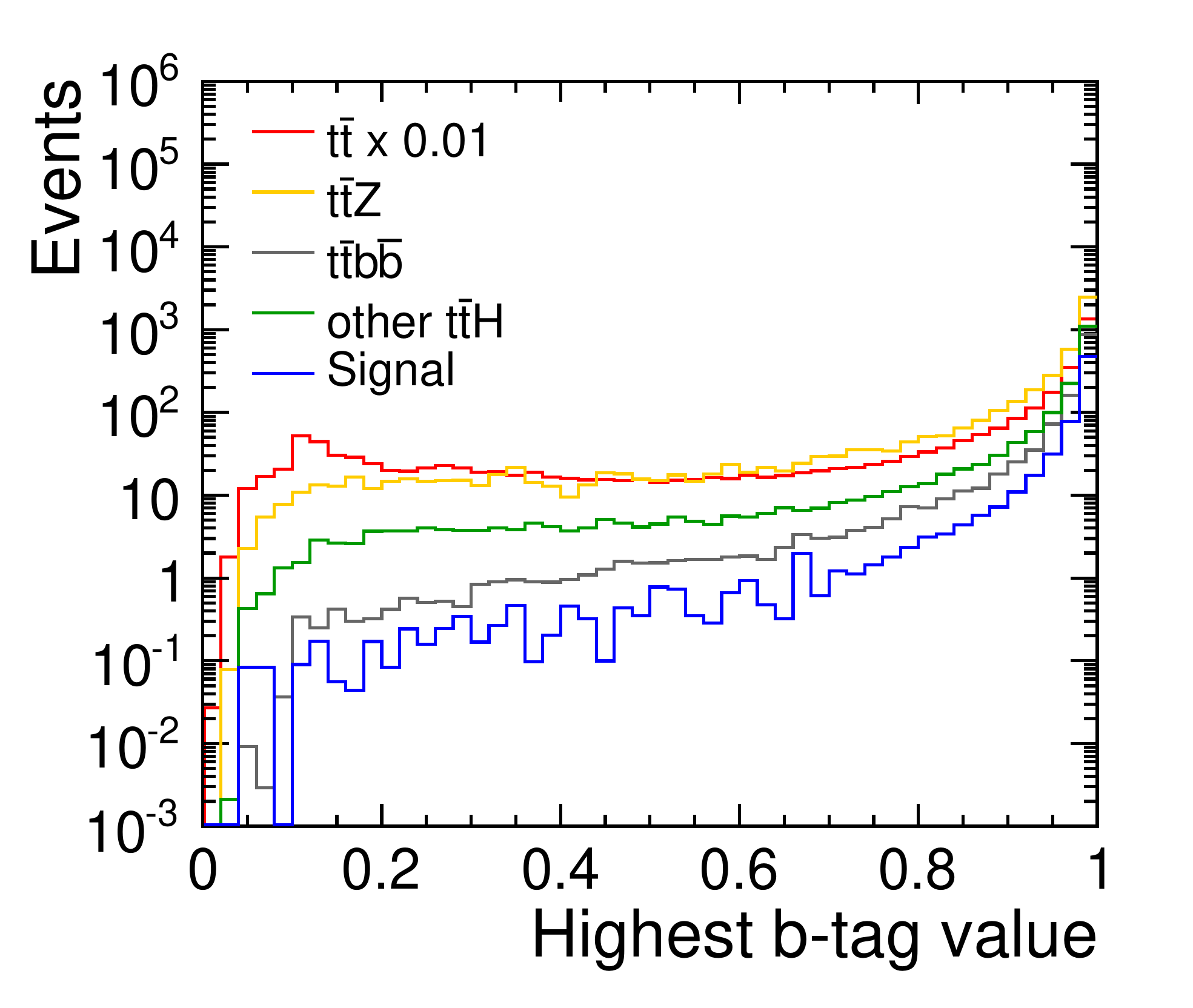} \hspace{0.5cm}
\includegraphics[width=0.45\textwidth]{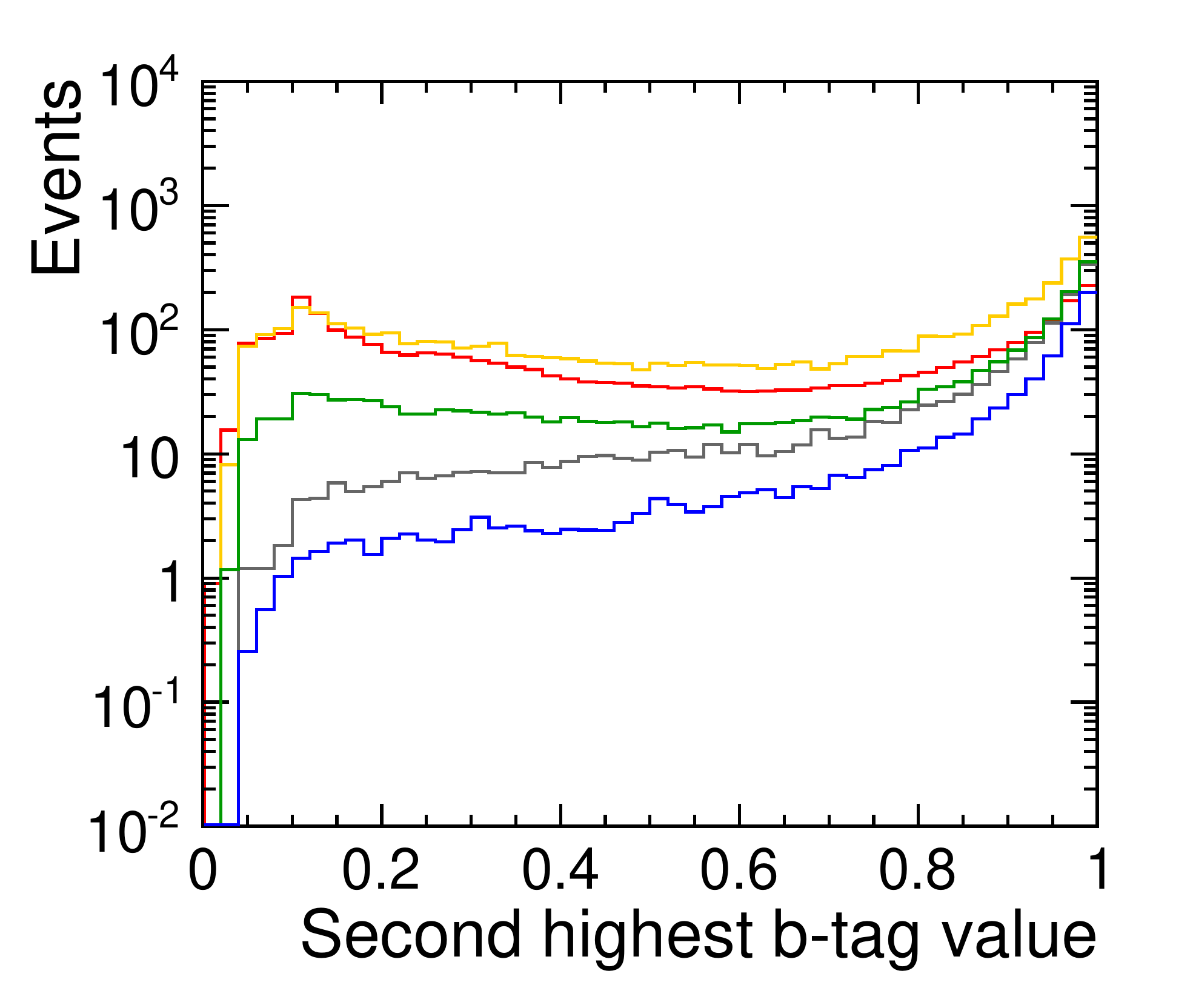} \\
\includegraphics[width=0.45\textwidth]{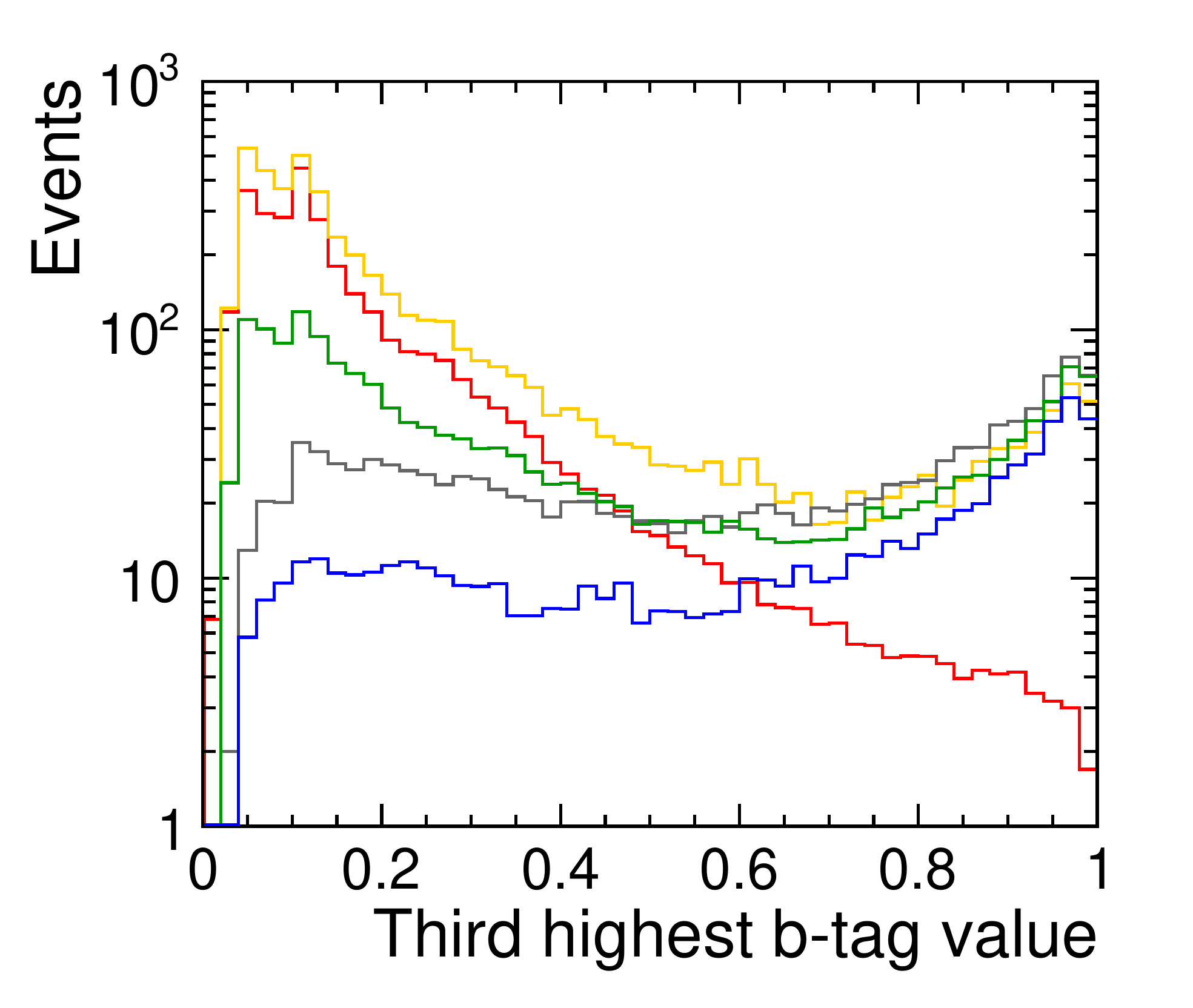} \hspace{0.5cm}
\includegraphics[width=0.45\textwidth]{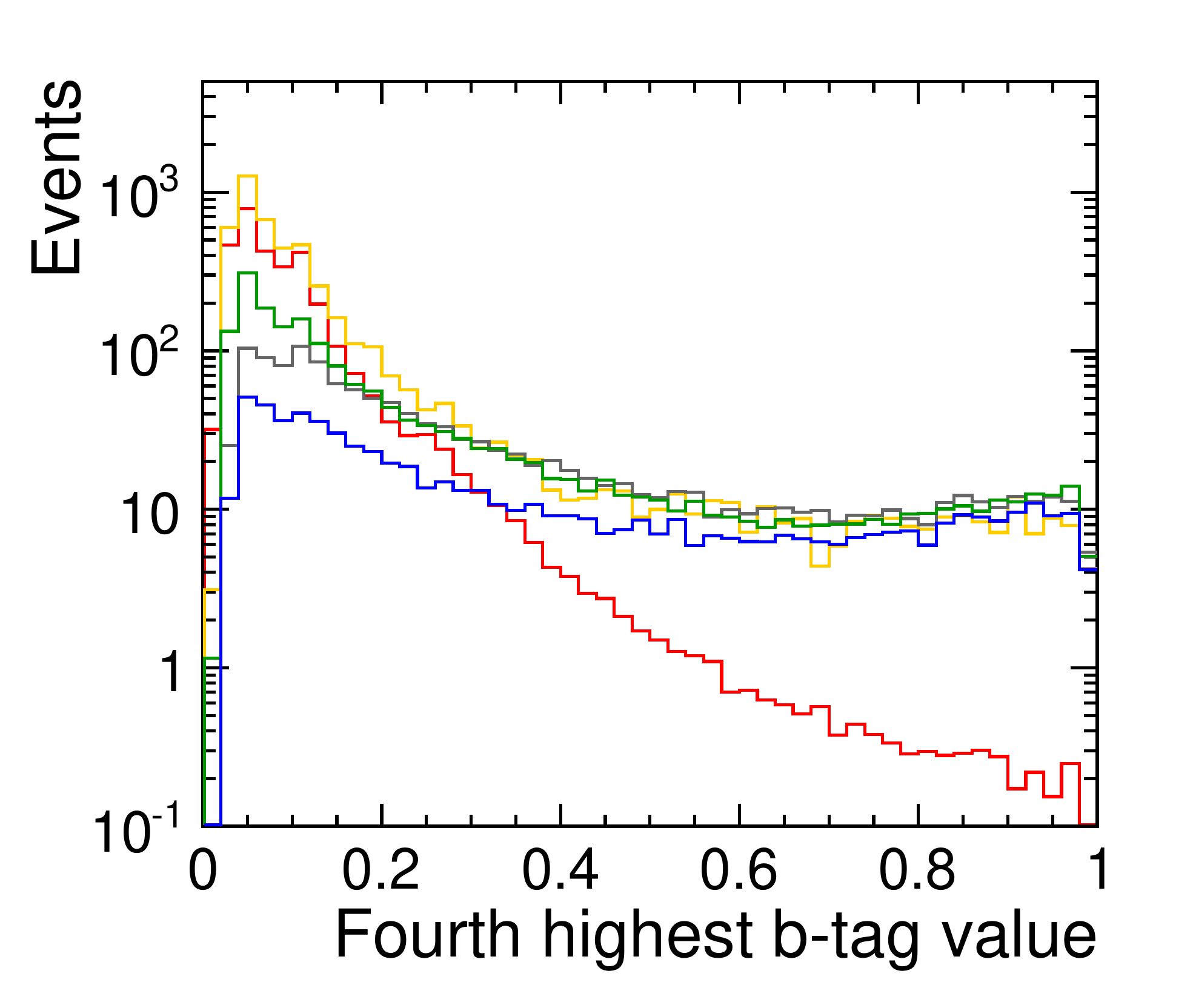} \\
\caption{\label{sid:benchmarking:fig:bt_outputs_tth_eightJet_btag} Distributions of several discriminating  
variables used in the event selection for the eight jet final state. The signals are shown in blue while the 
backgrounds are shown in different colours. The distribution for \toppair was scaled by a factor of $0.01$.}
\end{figure}

\begin{figure}[H]
\includegraphics[width=0.45\textwidth]{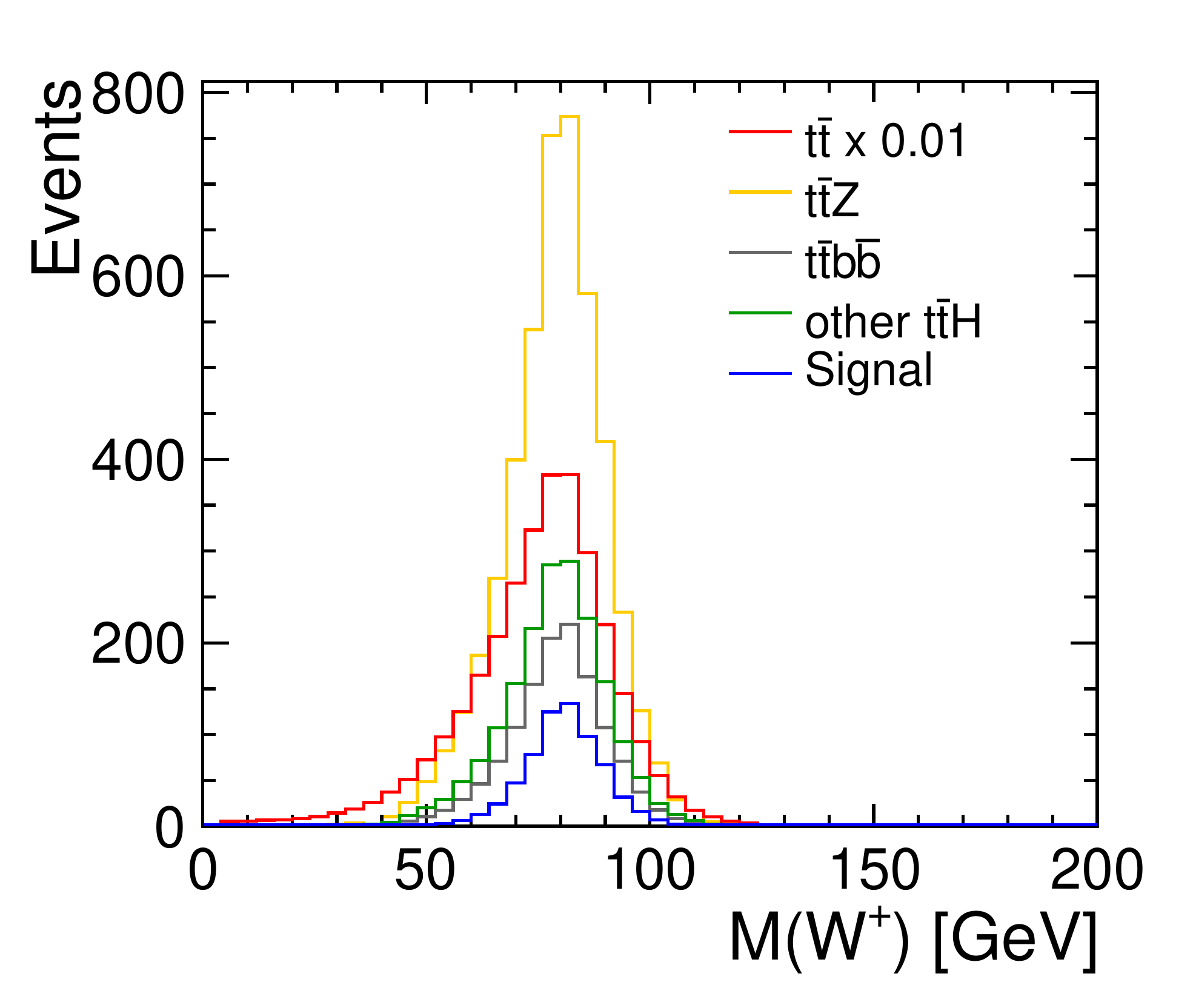} \hspace{0.5cm}
\includegraphics[width=0.45\textwidth]{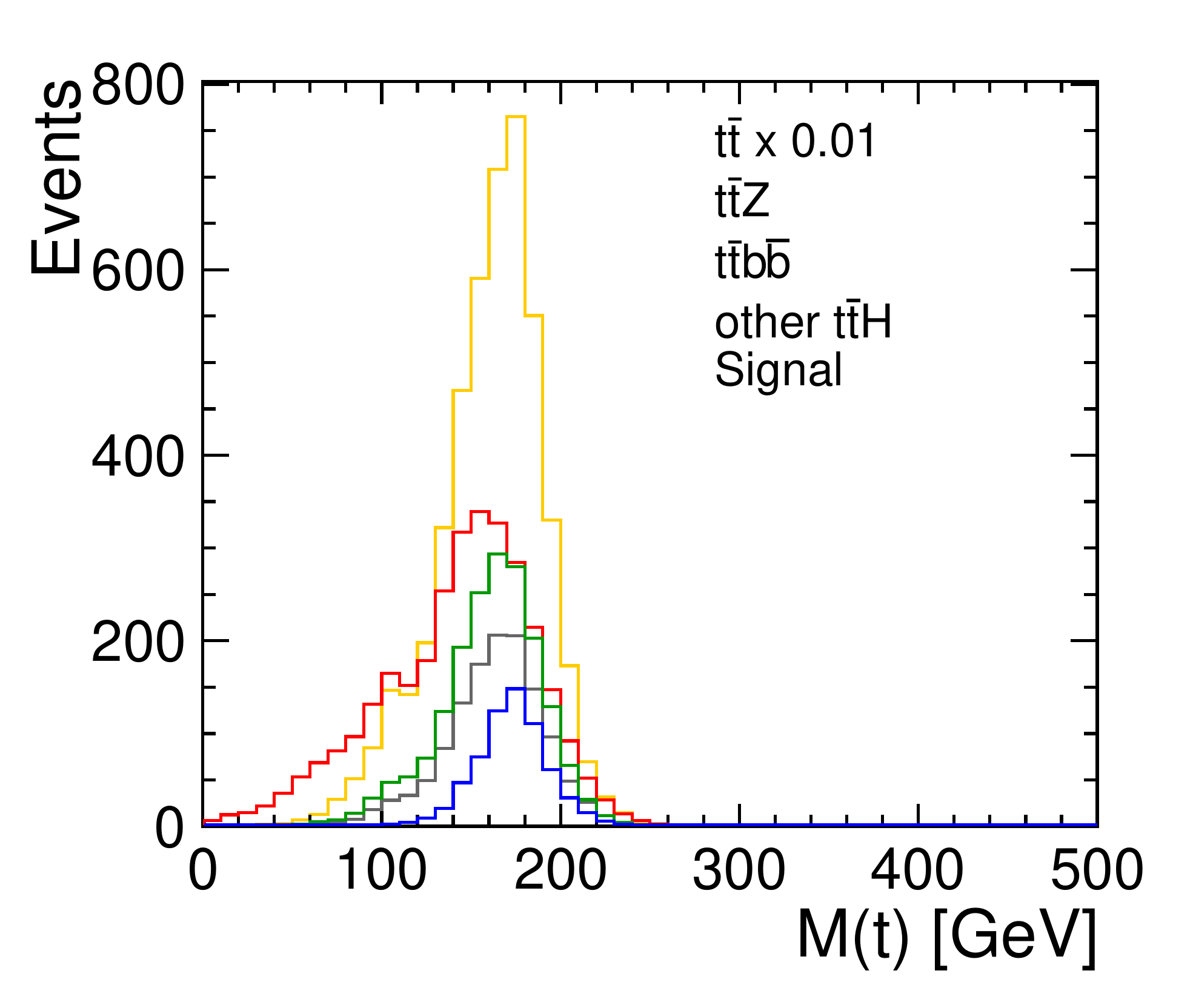} \\
\includegraphics[width=0.45\textwidth]{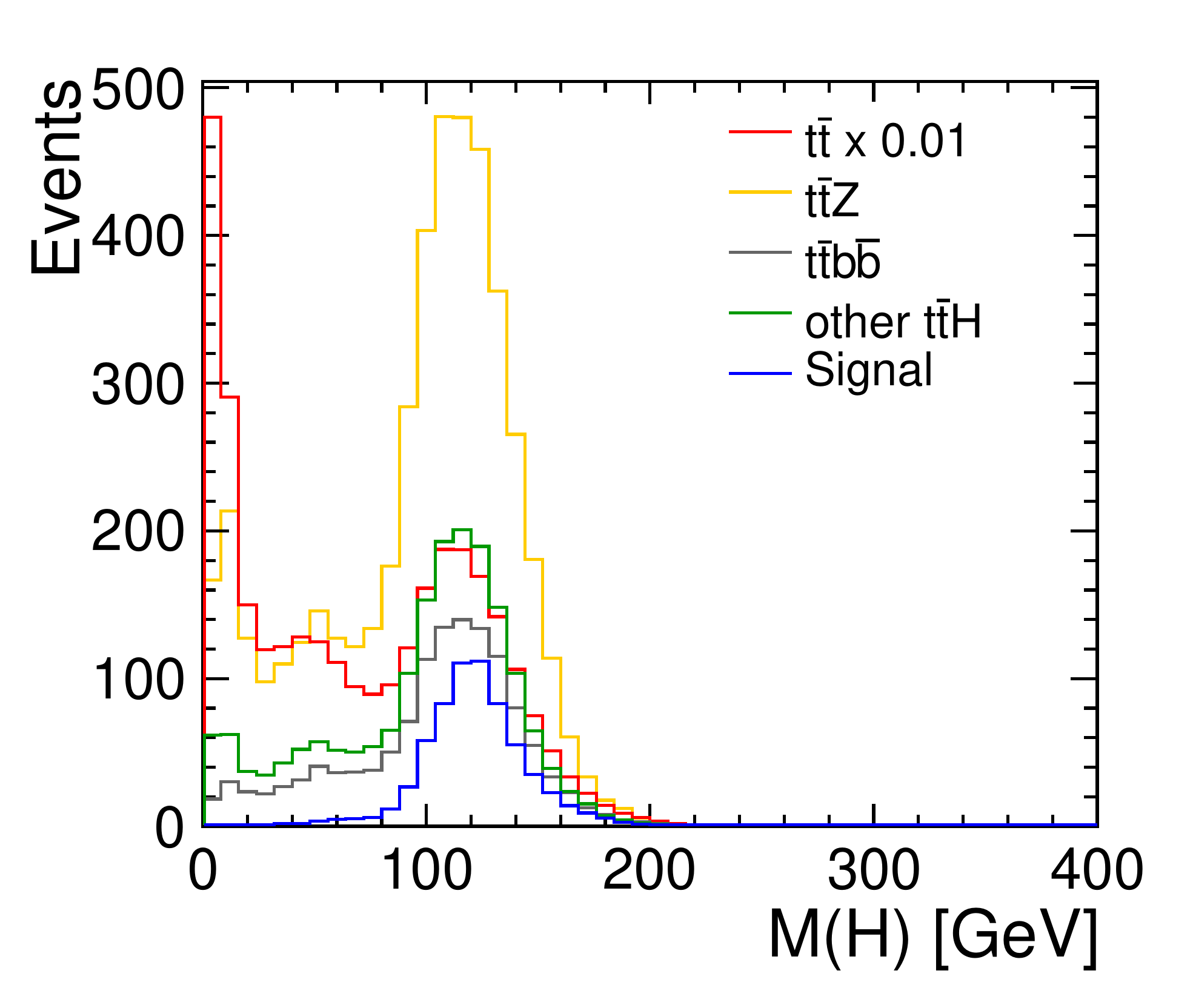} \hspace{0.5cm}
\includegraphics[width=0.45\textwidth]{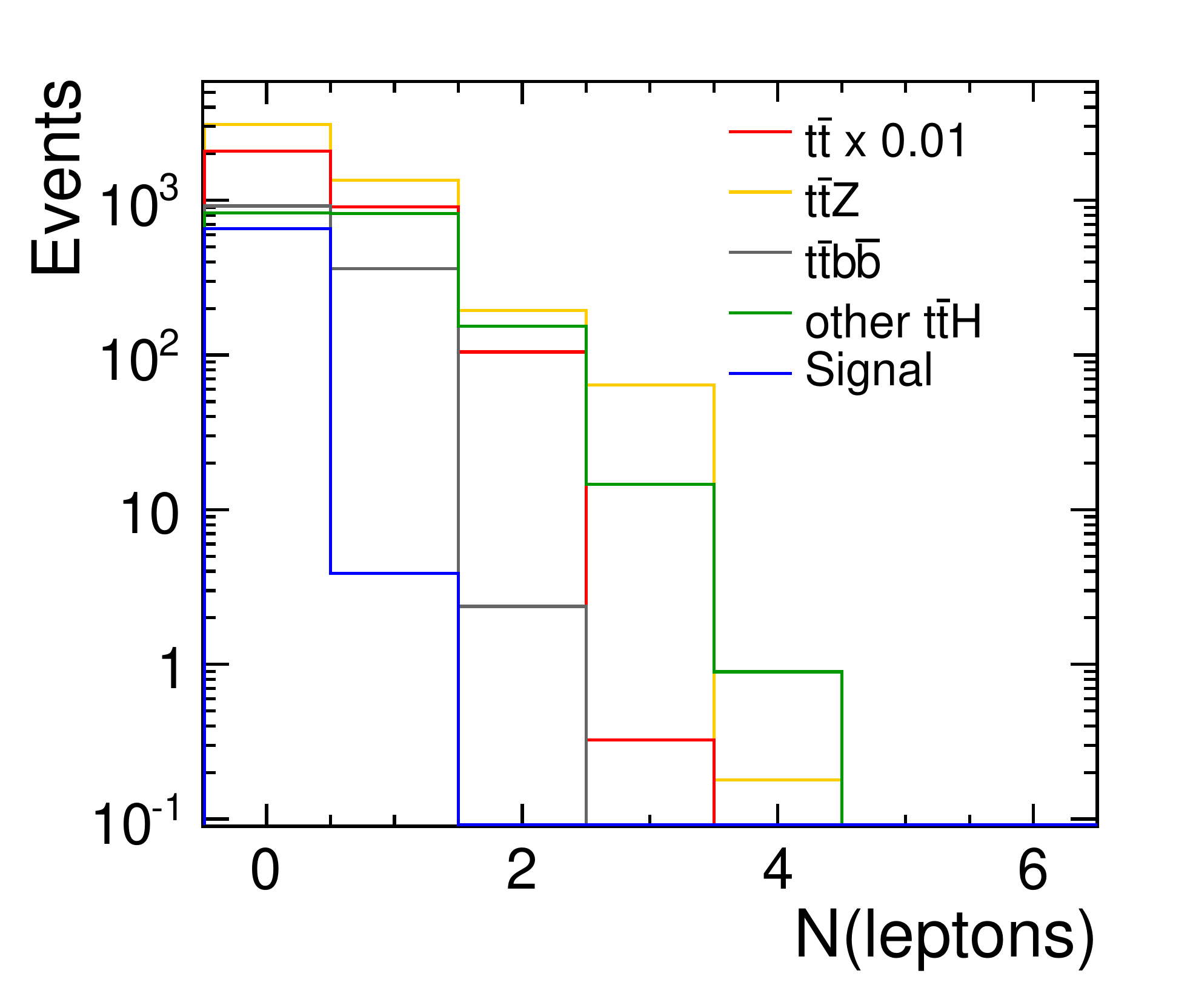} \\
\includegraphics[width=0.45\textwidth]{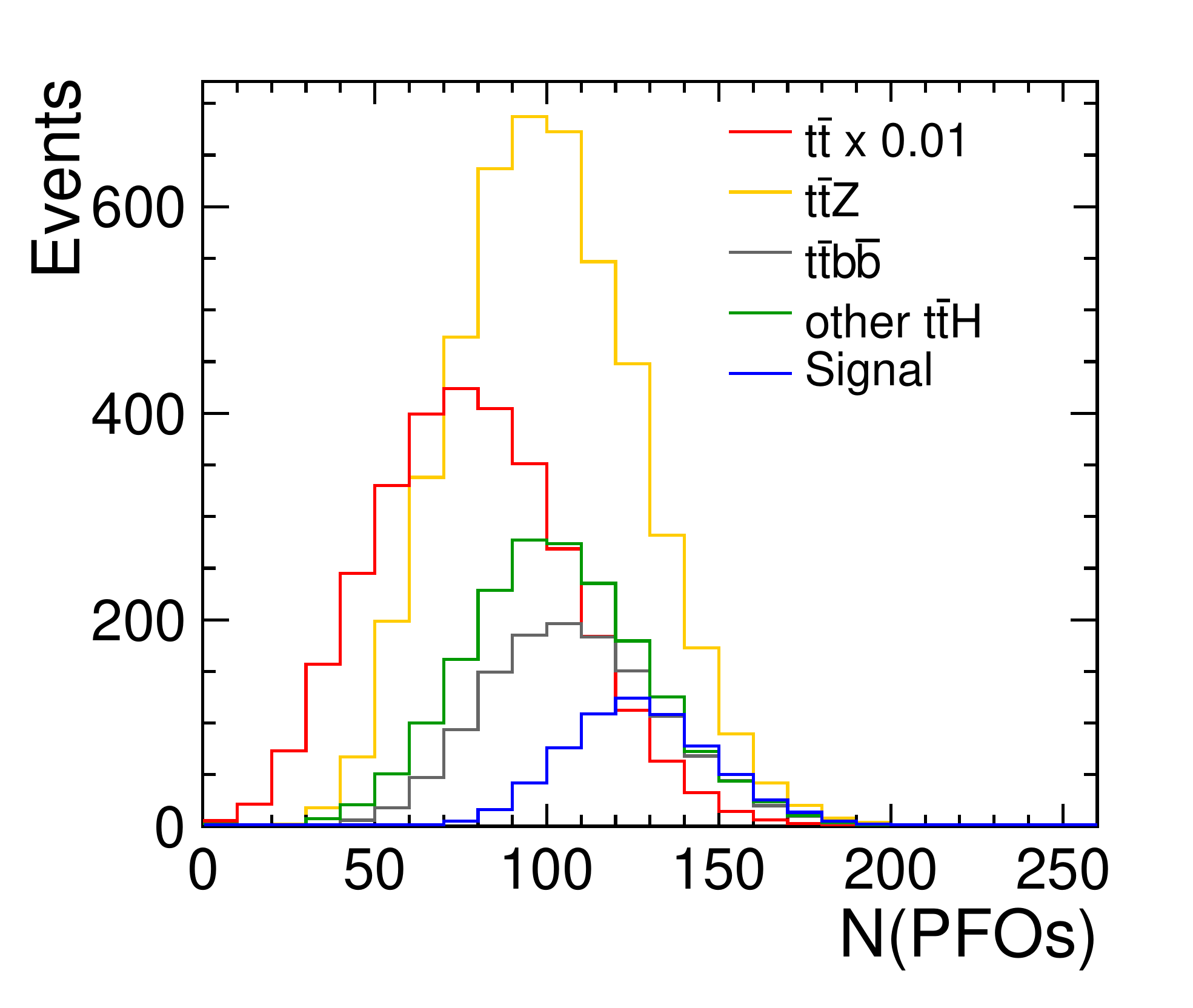} \hspace{0.5cm}
\includegraphics[width=0.45\textwidth]{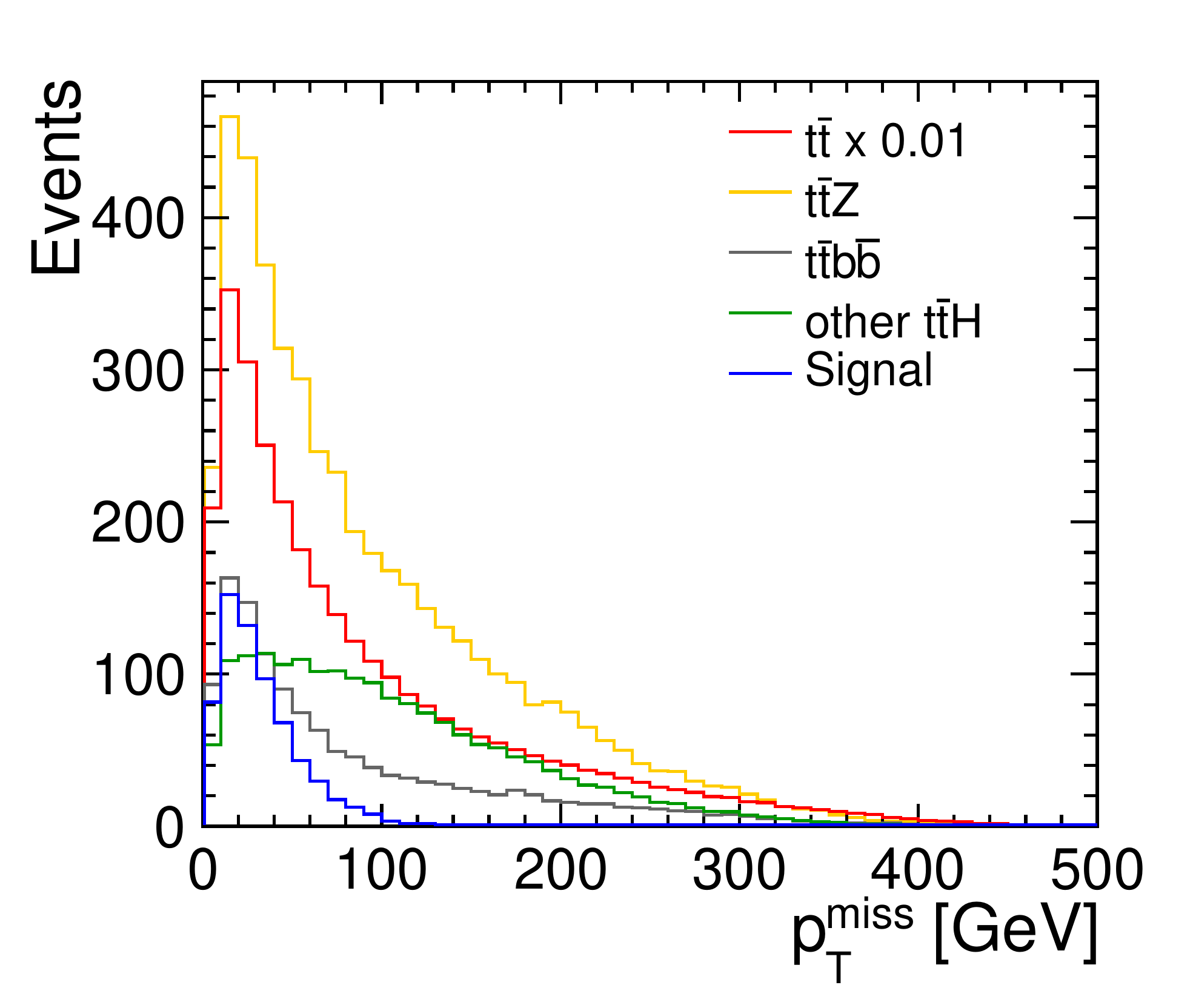} \\
\caption{\label{sid:benchmarking:fig:bt_outputs_tth_eightJet_mass} Distributions of several discriminating  
variables used in the event selection for the eight jet final state. The signals are shown in blue while the 
backgrounds are shown in different colours. The distribution for \toppair was scaled by a factor of $0.01$.}
\end{figure}

\begin{figure}[H]
\includegraphics[width=0.45\textwidth]{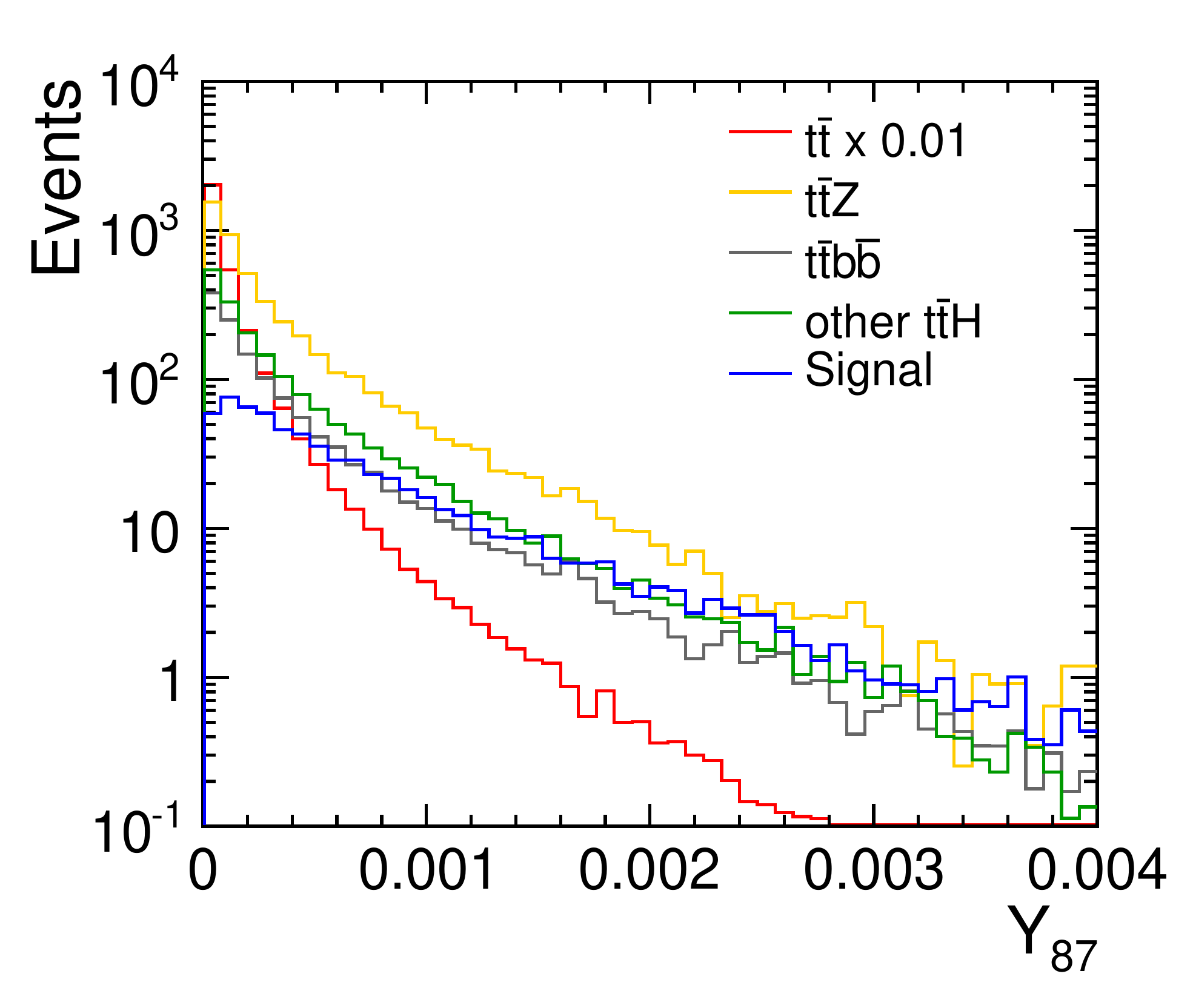} \hspace{0.5cm}
\includegraphics[width=0.45\textwidth]{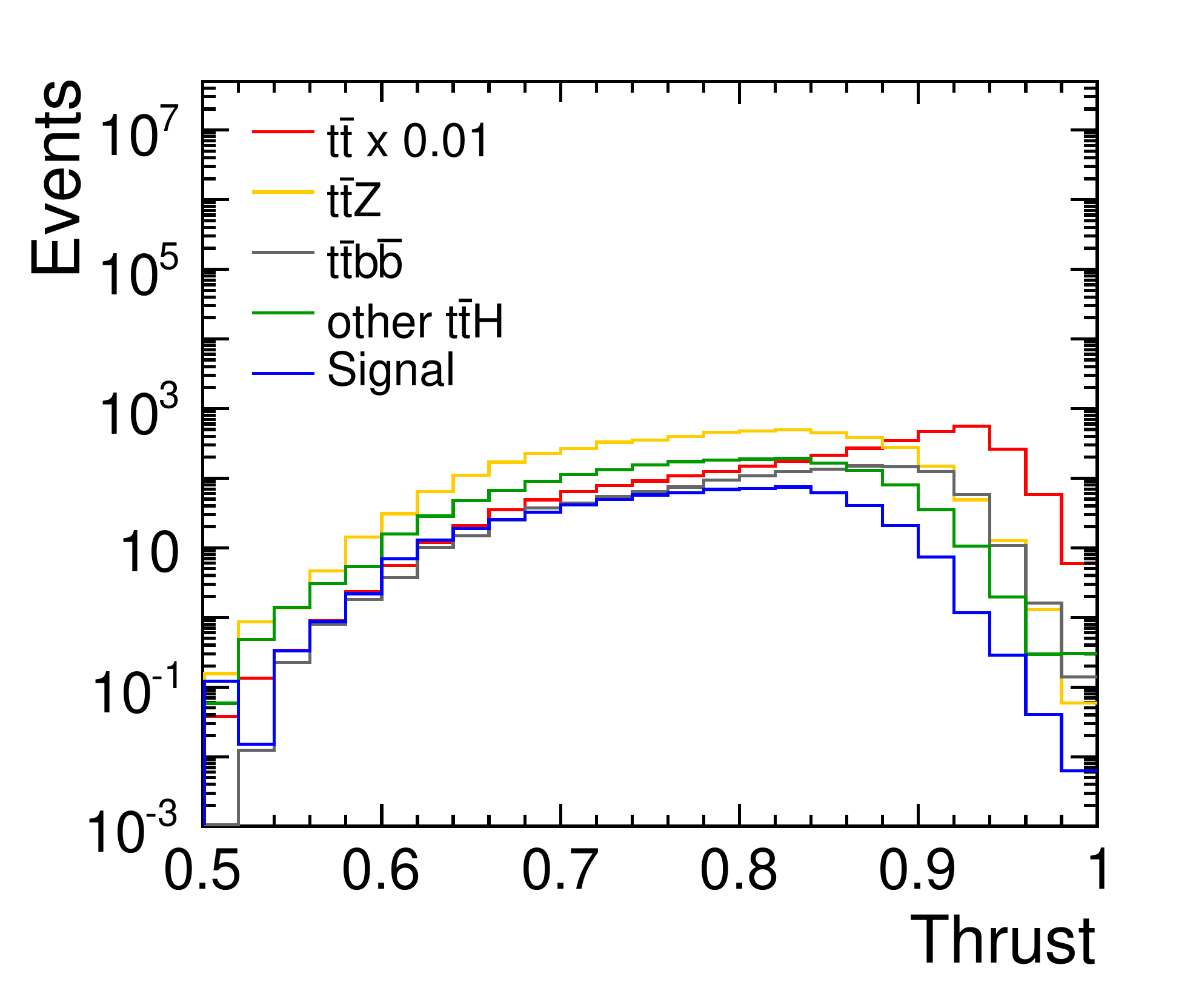} \\
\includegraphics[width=0.45\textwidth]{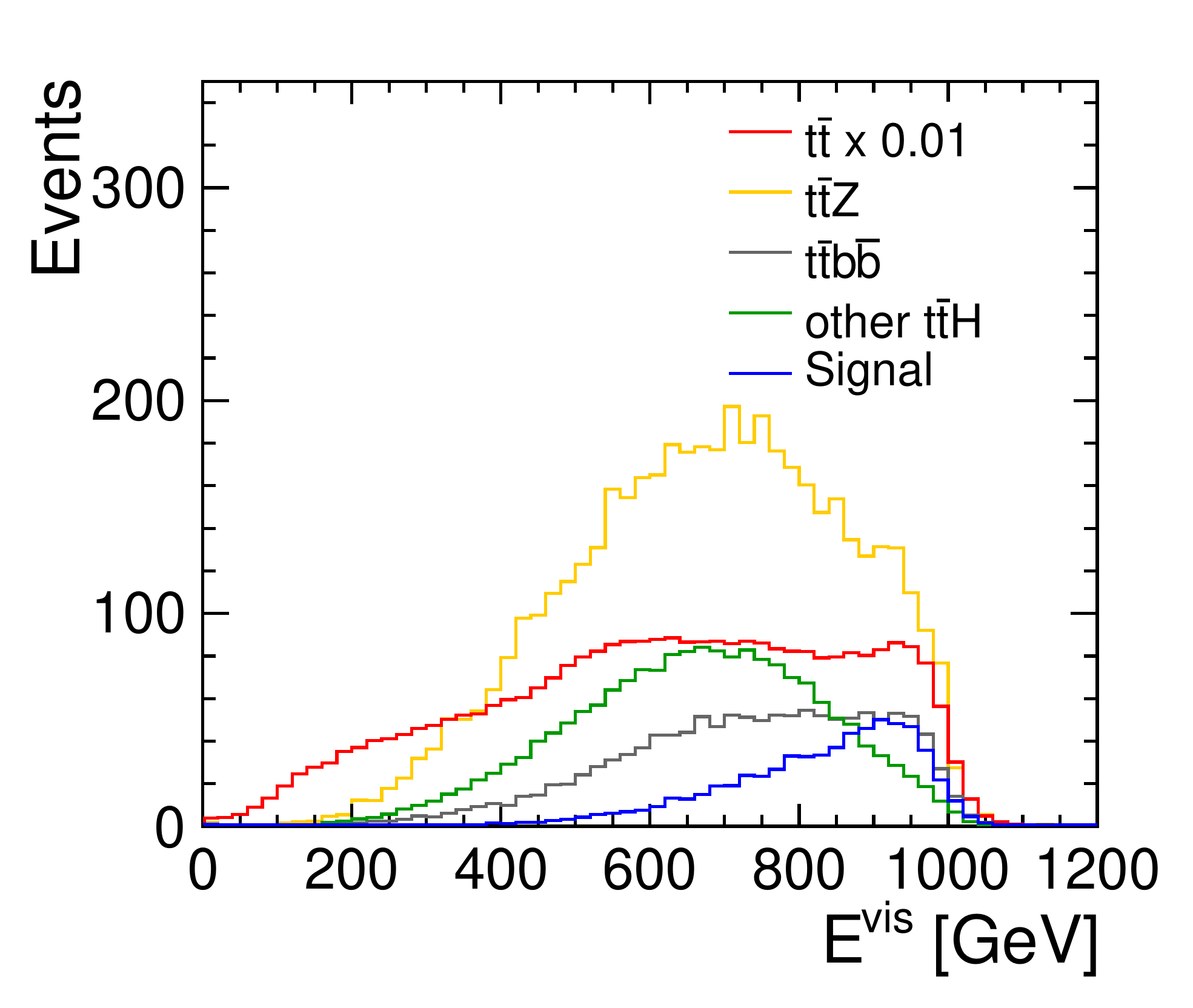} \hspace{0.5cm}
\caption{\label{sid:benchmarking:fig:bt_outputs_tth_eightJet_eventShape} Distributions of several discriminating  
variables used in the event selection for the eight jet final state. The signals are shown in blue while the 
backgrounds are shown in different colours. The distribution for \toppair was scaled by a factor of $0.01$.}
\end{figure}

\newpage

\section{Number of selected events with preselection on the number of isolated leptons}
\label{sec:appendix_selected_events_preselection}

\begin{table}[H]
\caption{\label{tab:tth_selected_events_preselection} Number of selected events for the different 
final states assuming an integrated luminosity of 1~\abinv. The values obtained for the six 
and eight jet final state selections are shown separately. Events with one isolated lepton are preselected for the 
six jet final state and events without isolated leptons are preselected for the eight jet final state before 
the training of the BDTs.}
\begin{center}
\begin{tabular}{l R[.]{3}{1} R[.]{3}{1}} \toprule
Final state & \multicolumn{1}{c}{BDT trained to select 6 jets} & \multicolumn{1}{c}{BDT trained to select 8 jets} \\ \midrule
\ttH, $\smH \to \bpair$ (6 jets) & 191.6 & 57.4 \\
\ttH, $\smH \to \bpair$ (8 jets) & 1.6 & 299.4 \\
\ttH, $\smH$ not $\bpair$ (6 jets) & 9.6 & 2.8 \\
\ttH, $\smH$ not $\bpair$ (8 jets) & 2.5 & 12.4 \\
\ttH (4 jets) & 20.9 & 1.4 \\
$\toppair Z$ & 105.6 & 187.1 \\
$\toppair g^{*} \to \toppair\bpair$ & 100.1 & 180.7 \\
$\toppair$ & 232.0 & 381.6 \\ \bottomrule
\end{tabular}
\end{center}
\end{table}

\newpage

\bibliographystyle{cliccdr}
\bibliography{lcd}

\end{document}